%% file: CEC_HetStorage_v5.tex
\documentclass[12pt,draftclsnofoot,peerreviewca,letterpaper,onecolumn]{IEEEtran}

\usepackage{cite}
\usepackage{graphicx,color,epsfig,rotating}
\usepackage{amsfonts,amsmath,amssymb,bbm}
\usepackage{algorithm}
\usepackage{algpseudocode}
\usepackage{subfigure}
\usepackage{amsmath}
\usepackage{cite}
\usepackage{mdwtab}
\usepackage{subfigure}
\usepackage{placeins}
\usepackage{psfrag, graphicx}
\usepackage[latin1]{inputenc}
\usepackage{amssymb}
\usepackage{makeidx}
\usepackage{epstopdf}
\usepackage{enumitem}
\usepackage{amssymb}

\input{macros}

\newtheorem{defn}{Definition}
\newtheorem{theorem}{Theorem}
\newtheorem{lemma}{Lemma}
\newtheorem{corollary}{Corollary}
\newtheorem{claim}{Claim}
\newtheorem{remark}{Remark}

\IEEEoverridecommandlockouts

\begin{document}
\setcounter{page}{1}

\title{Coded Elastic Computing on Machines with Heterogeneous Storage and Computation Speed}
\author{Nicholas Woolsey,~\IEEEmembership{Student Member,~IEEE}, Rong-Rong Chen,~\IEEEmembership{Member,~IEEE},\\ and Mingyue Ji,~\IEEEmembership{Member,~IEEE}
\thanks{This manuscript was partially presented in the conference paper \cite{woolsey2019hetCEC}.}
\thanks{The authors are with the Department of Electrical Engineering,
University of Utah, Salt Lake City, UT 84112, USA. (e-mail: nicholas.woolsey@utah.edu, rchen@ece.utah.edu and mingyue.ji@utah.edu)}
}

\maketitle

\begin{abstract}
We study the optimal design of heterogeneous Coded Elastic Computing (CEC) where machines have varying computation speeds and storage. CEC introduced by Yang et al. in 2018 is a framework that mitigates the impact of elastic events, where machines can join and leave at arbitrary times. In CEC, data is distributed among machines using a Maximum Distance Separable (MDS) code such that subsets of machines can perform the desired computations. However, state-of-the-art CEC designs only operate on homogeneous networks where machines have the same speeds and storage. This may not be practical. In this work, based on an MDS storage assignment, we develop a novel computation assignment approach for heterogeneous CEC networks to minimize the overall computation time. We first consider the scenario where machines have heterogeneous computing speeds but same storage and then the scenario where both heterogeneities are present. We propose a novel combinatorial optimization formulation and solve it exactly by decomposing it into a convex optimization problem for finding the optimal computation load and a ``filling problem" for finding the exact computation assignment. A low-complexity ``filling algorithm" is adapted and can be completed within a number of iterations equals at most the number of available machines. 
\if{0}
 We study the optimal design of  heterogeneous Coded Elastic Computing (CEC) networks where machines have  varying properties of  computation speed and  storage capacity. CEC introduced by Yang {\it et al.} in 2018 is a framework that mitigates the impact of elastic events, where machines can join and leave the network at arbitrary times. In CEC, a set of data is distributed among storage constrained machines using a Maximum Distance Separable (MDS) code such that subsets of machines can perform the desired computations. This  eliminates the need to re-distribute data after each elastic event. However, state-of-the-art CEC designs only operate on homogeneous networks where machines have  the same computation speeds and storage capacity. This may not be the case in practice. In this work, based on an MDS like storage scheme for heterogeneous storage, we develop a novel approach for 
general heterogeneous CEC networks to minimize the overall computation time by distributing an optimal computation load or number of computations assigned to each machine.  We first focus on the scenario where machines  have heterogeneous computing speeds but the same storage capacity and then the scenario where both heterogeneities are present. We propose a novel problem formulation to minimize the overall computation time by solving a combinatorial optimization problem. We  solve this problem  exactly by decomposing it into two sub-problems including a relaxed convex optimization problem for finding the optimal computation load and a ``filling problem" for finding the exact computation assignment. 
A low complexity ``filling algorithm" is adapted to find the optimal computation assignment within a number of iterations equals at most the number of available machines. 
The proposed CEC computation assignment algorithm works for an arbitrary set of machine speeds and storage requirements.
\fi

\end{abstract}


\section{Introduction}
\label{section: intro}
Coding is an effective tool to speed up distributed computing networks and has attracted significant attention recently. Examples include Coded Distributed Computing (CDC) for MapReduce-like frameworks  \cite{li2018fundamental,li2018cdc,konstantinidis2018leveraging,woolsey2018new,Srinivasavaradhan2018distributed,prakash2018coded,woolsey2019ccdc,xu2019cdc,woolsey2019coded,wan2020topological}  
and coded data shuffling for distributed machine learning \cite{attia2019shuffling,elmahdy2018shuffling,wan2020fundamental}, where code designs minimize the communication load by trading increased computation resources and/or storage on each machine. Another example is to use codes to mitigate the straggler effect in applications such as linear operations 
\cite{lee2017speeding,tandon2017gradient,dutta2016short,wan2020distributed} 
and matrix multiplications \cite{yu2020straggler,dutta2020coding,wan2020cache}, where any subset of machines with a cardinality larger than the recovery threshold can recover the matrix multiplication. This eliminates the need to wait for the computation of slow machines. Moreover, coding has also been applied to the computing problems integrated with optimizations \cite{karakus2017straggler} and with the requirements of security \cite{bitar2017minimizing,aliasgari2020computing}, privacy \cite{hua2019privacy,obead2020private,aliasgari2020computing,chen2020gcsa} and robustness \cite{chen2018draco}. In this paper, we will consider another important and novel application of coding in order to cope with the {\em elasticity} in distributed and cloud computing systems \cite{jonas2017occupy}.

Coded Elastic Computing (CEC) was introduced by Yang {\it et al.} in 2018 to mitigate the impact of {\it preempted} machines on a storage limited distributed computing network \cite{yang2018coded}. As opposed to stragglers, whose identities are unknown when computations are assigned, the preempted (unavailable) machines are known. Hence, while it is necessary to bring computation redundancy into the straggler mitigation problem, it is desired to have no computation redundancy in the elastic computing problem.  In elastic computing, computations are performed over many times steps and between each time step an {\it elastic event} may occur where machines become preempted or available again. Computations are performed on the same set of data, e.g., a matrix, while the computations change each time step. For example, in each time step the data matrix may be multiplied with a different vector. In each time step, the goal becomes to assign computations among the available machines such that the overall computation time can be minimized. A naive approach is to assign each machine a non-overlapping part of the data. However, this is inefficient as the storage placement has to be redefined with each elastic event. 

In homogeneous CEC proposed in \cite{yang2018coded}, where all machines have the same storage space and computing speed, the storage of each machine is placed once using a Maximum Distance Separable (MDS) code and remains unchanged between elastic events. The data is split into $L$ equal sized, disjoint data sets and each machine stores a coded combination of these sets. For the state-of-the-art design \cite{yang2018coded}, each machine stores 
a unique and coded $\frac{1}{L}$ fraction (in size) of the original data library. Furthermore, since all the machines have the same computing speed, in order to minimize the overall computation time, 
all the machines are assigned an equal number of computations (e.g., number of vector-vector multiplications). The requirement for CEC is to allow any computation task (e.g., matrix-vector multiplications) be resolved by combining the coded computation results of $L$ machines.
This means that the computation tasks have to be assigned to these $L$ available machines while keeping the computation redundancy to a minimum.
The authors of \cite{yang2018coded} proposed a novel ``cyclic'' computation assignment such that each machine is assigned by the same number computations and no computation redundancy is present.

The recent work \cite{dau2019optimizing} also studies the homogeneous CEC and aims to maximize the overlap of the task assignments between computation time steps. With each elastic event, the computation assignment must change. In the cyclic approach proposed in \cite{yang2018coded}, the assignments in the current time step are independent of assignments in previous time steps. In \cite{dau2019optimizing}, the authors design assignment schemes to minimize the changes in the assignments between time steps. In some cases, the proposed assignment schemes were shown to achieve zero ``transition waste", or 
no new local computations at the existing machines.

The works of \cite{yang2018coded} and \cite{dau2019optimizing} only study homogeneous CEC networks. However, in practice, the available computing machines in a cloud system are often heterogeneous due to the fact that even if all the machines are homogeneous, one machine could be used by multiple users simultaneously and each user may have different computation and storage demands.  When machines have varying storage space and computing speeds or capabilities, the previous CEC designs are sub-optimal. For example, in \cite{yang2018coded} and \cite{dau2019optimizing}, each machine is assigned the same amount of computations. If one machine is faster it will be idle waiting for slower machines to finish. The problem of heterogeneous computation assignment is challenging because we must ensure that each computation is performed on $L$ coded data sets while meeting some optimal computation load of each machine based on the relative computing speeds of the machines such that the overall computation time can be minimized. Moreover,  the previous designs assume each machine is capable of storing a $\frac{1}{L}$ fraction of the data, or one coded data set. If machines have varying storage requirements then to use the designs of \cite{yang2018coded} and \cite{dau2019optimizing}, either machines with less storage are excluded or machines with more storage do not utilize their entire storage.


In this paper, we propose a CEC framework optimized for a heterogeneous network where machines have varying computation speeds and storage requirements such that the overall computation time is minimized. 
{We represent the relative speed of the machines  by a speed vector $\boldsymbol{s}\in\mathbb{R}_{+}^N$ and the storage capacity of the machines by a storage vector $\boldsymbol{\sigma} \in \mathbb{Z}_+^N$. The latter represents an integer number of coded data sets each machine can store and imposes a limit on the amount of computations each machine can be assigned. Under this framework, as much as the storage allows, more computations are assigned to faster machines and less to slower machines in a systematic way to minimize the 
overall computation time while maintaining MDS code requirements.
}  
The main contributions of this paper are summarized as follows.
\begin{itemize}

\item  To the best of our knowledge, this is the first work to adopt the performance metric of overall computation  time to optimize the computation assignment for CEC networks while making use of MDS codes. 
We introduce a class  of computation assignments for CEC networks in which computations are distributed
over common row sets ({with possibly different sizes}) across all machines, and the optimization of computation assignment is performed through an iterative  algorithm which identifies the row sets and the machines that compute them. These lead to a novel combinatorial optimization framework  
that finds
the optimal assignment for minimal overall computation  time.  

\item   As opposed to other existing works on CEC networks that focus on homogeneous networks, our study applies to general CEC networks in which machines can have both  heterogeneous computation speed and storage capacity.  We develop a novel approach to solve the optimal computation assignment by utilizing both types of heterogeneity under a unified optimization framework. Particularly, we show that the optimal computation assignment can be determined based on an ordering of each machine's storage capacity to computation speed ratio (SCR). The overall computation  time of the optimal computation assignment is limited by machines with the largest SCR.
\item We  solve the proposed combinatorial optimization problem exactly by decomposing it into two sub-problems. The first is a relaxed convex optimization problem to determine the computation load of each machine {without specifying the exact computation assignment}. The second is a computation assignment problem (or a ``filling problem") to find the exact computation assignment across machines while meeting the optimal computation load {obtained from the first sub-problem} and the MDS code requirement. 
    \item 
    Under the proposed optimization framework,  after solving the first sub-problem for the optimal computation load, we can adapt an iterative filling algorithm (previously developed for setting of private information retrieval  \cite{woolsey2019optimal}) to find the exact computation assignment in the second sub-problem.
    The adapted algorithm for the new setting of CEC networks 
    converges within a number of iterations no greater than the number of available machines. This leads to a low-complexity design of  optimal computation assignment for large CEC networks with an arbitrary set of machine computing speeds and storage requirements. 

\end{itemize}

The paper is organized as follows. In Section~\ref{sec: Network Model and Problem Formulation}, we present the network model and introduce the proposed problem formulation. Before going into details of the proposed solutions, in Section~\ref{sec: example}, we give examples of the proposed CEC algorithms for two scenarios, which are networks with 1) heterogeneous computing speeds and homogeneous storage space and 2) heterogeneous computing speeds and heterogeneous storage capacity. The proposed combinatorial optimization problem is solved in two steps in Section~\ref{sec: compload} and Section~\ref{sec: compassign}, where the proposed low-complexity CEC computation assignment algorithms are introduced. The paper is concluded in Section~\ref{sec: conclusions}.

\paragraph*{Notation Convention}
We use $|\cdot|$ to denote the cardinality of a set or the length of a vector. 
Let $[n] := \{1,2,\ldots,n\}$ denote a set of integers from $1$ to $n$. A bold symbol such as $\boldsymbol{a}$ indicates a vector and $a[i]$ denotes the $i$-th element of $\boldsymbol{a}$. 
$\mathbb{Z}^+$ denotes the set of all positive integers; $\mathbb{R}^+$ denotes the set of all positive real numbers and $\mathbb{Q}^+$ denotes the set of all positive rational numbers. Finally, let $\mathbb{R}_+^N$ be the set of all length-$N$ vectors of real, positive numbers.

\section{System Model and Problem Formulation}
\label{sec: Network Model and Problem Formulation}

\subsection{System Model }

We consider a set of $N$ machines. Each machine $n\in[N]$ stores an integer number, $\sigma[n]$, of coded sub-matrices, which we refer to as {cs-matrices}, derived from a $q\times r$ data matrix, $\boldsymbol{X}$, where $q$ can be large. Here, we denote the vector $\boldsymbol{\sigma} = (\sigma[1], \sigma[2], \ldots, \sigma[N]), \;\sigma[n] \in \mathbb{Z}^+,\; \forall n \in [N]$ as the storage vector and 
\be
Z\triangleq\sum_{n=1}^{N}\sigma[n]
\ee
is the total number of cs-matrices stored among all machines.\footnote{Note that assuming  $\sigma[n] \in \mathbb{Z}^+, \forall n \in [N]$ is for the ease of presentation. If any $\sigma[n]$ is a fractional number, then we can let $Z$ be large such that all $\sigma[n]$s can be a positive integer.} The cs-matrices are specified by an $Z\times L$ MDS code generator matrix $\boldsymbol{G}$ where $g_{i,\ell}$ denotes the element in the $i$-th row and $l$-th column, where $Z,L \in \mathbb{Z}^+$. 
Let any $L$ rows of $\boldsymbol{G}$ be invertible. The data matrix, $\boldsymbol{X}$, is row-wise split into $L$ disjoint, $\frac{q}{L}\times r$ uncoded sub-matrices, $\boldsymbol{X}_1,\ldots ,\boldsymbol{X}_L$. Define a set of $Z$ cs-matrices
\be
\boldsymbol{\tilde{X}}_i = \sum_{\ell=1}^{L}g_{i,\ell}\boldsymbol{X}_\ell
\ee
for $i\in[Z]$. Each  $ \boldsymbol{\tilde{X}}_i, i \in [Z]$ has $\frac{q}{L} $ rows.
Assume that machine $n$ will store $\sigma[n]$ of these cs-matrices, specified by an index set $\mathcal{Q}_n$.
In other words, machine $n$ stores the cs-matrices of $\{\boldsymbol{\tilde{X}}_i: i\in\mathcal{Q}_n\}$, where $|\mathcal{Q}_n|=\sigma[n]$.
We also assume that different machines store different cs-matrices: the sets $\mathcal{Q}_1, \cdots, \mathcal{Q}_N$ are disjoint.
Also, note that given a cs-matrix $ \boldsymbol{\tilde{X}}_i, i \in [Z]$, there is an unique machine that stores this cs-matrix.

The machines collectively perform matrix-vector computations\footnote{It can be shown that our CEC designs also operate on the other applications outlined in \cite{yang2018codedArxiv} rather than just matrix-vector multiplications including matrix-matrix multiplications, linear regressions and so on.} over multiple times steps. In a given time step only a subset of the $N$ machines are available to perform matrix computations. Note that, the available machines of each time step are known when we design the computation assignments. Specifically, in time step $t$, a set of available machines $\mathcal{N}_t \subseteq [N]$ aims to compute
\be
\boldsymbol{y}_t = \boldsymbol{X}\boldsymbol{w}_t
\ee
where $\boldsymbol{w}_t$ is some vector of length $r$. The machines of $[N]\setminus\mathcal{N}_t$ are preempted. The number of cs-matrices stored at the available machines is
\be
Z_t\triangleq\sum_{n\in\mathcal{N}_t}\sigma[n] \geq L,
\ee
which means that $Z_t$ is at least $L$ such that the desired computation can be recovered.


Machines of $\mathcal{N}_t$ do not compute $\boldsymbol{y}_t$ directly. Instead, each machine $n\in \mathcal{N}_t$ computes the set
\be
\mathcal{V}_{n} = \bigcup_{i\in\mathcal{Q}_n}\left\{ v = \boldsymbol{\tilde{X}}_i^{(j)}\boldsymbol{w}_t : j \in \mathcal{W}_{i} \right\}
\ee
where $\boldsymbol{\tilde{X}}_i^{(j)}$ is the $j$-th row of $\boldsymbol{\tilde{X}}_i$, the cs-matrix stored by machine $n$,  and $\mathcal{W}_{i}\subseteq\left[ \frac{q}{L}\right]$ is the set of rows of $\boldsymbol{\tilde{X}}_i$ assigned to machine $n$ in time step $t$ for computing tasks.
\subsection{Key Definitions }
In the following, we introduce three key definitions  that are useful to specify the computation assignments in a CDC network.
\begin{defn}
The  computation load vector, $\boldsymbol{\mu}$,  defined as
\be \label{eq: compload_vector}
\mu[n] = {\frac{|\mathcal{V}_n|}{\left( \frac{q}{L}\right)}=} \frac{\sum_{i\in\mathcal{Q}_n}|\mathcal{W}_{i}|}{
\left( \frac{q}{L} \right)}, \;\; \forall n \in \mathcal{N}_t,
\ee
is the normalized number of rows computed by  machine $n$ in time step $t$.
\hfill $\Diamond$
 \end{defn}
 
Note that while $\boldsymbol{\mu}$, $\mathcal{W}_{i}$, and  $\mathcal{V}_{n}$ can change with each time step,  reference to $t$ is omitted {for ease of presentation.  } Moreover, assume that  machines have varying computation speeds defined by a strictly positive vector $\boldsymbol{s} = [s[1], s[2], \ldots, s[N]], \;s[n] \in  \mathbb{Q}^+,\; \forall n \in [N]$,  fixed over all time steps. Here, computation speed is defined as the number of row multiplications per unit time.

 \begin{defn}
 Given a computation load vector  $\boldsymbol{\mu}$, the overall computation time is dictated by the machine(s) that takes the most time to perform its assigned computations. Thus, we define the overall computation  time  as a function of the computation load  $\boldsymbol{\mu}$,  as
\be
c(\boldsymbol{\mu}) = \max_{n\in \mathcal{N}_t} \frac{\mu[n]}{s[n]}. 
\ee
\hfill $\Diamond$
 \end{defn}

 At each time step $t$, for each $j\in\left[ \frac{q}{L}\right]$, the $j$-th row of $L$ sc-matrices undergoes a vector-vector multiplication with  $\boldsymbol{w}_t$. The results are sent to a master node which can resolve the elements of $\boldsymbol{y}_t$ by the MDS code design. To ensure each row is assigned $L$ times,\footnote{When each machine stores just $1$ coded matrix, we say the computations are assigned to each machine as in the original CEC work \cite{yang2018codedArxiv}. Alternatively, when some machines store more than $1$ matrix, each row is computed for $L$ different matrices which are stored across a number of machines less than or equal to $L$.} we will introduce a general framework for defining computation assignments. 

Given a cs-matrix $\boldsymbol{\tilde{X}}_i$,  $\mathcal{W}_{i}$ includes the subset of rows in  $\boldsymbol{\tilde{X}}_i$ {assigned to be computed by the machine that stores $\boldsymbol{\tilde{X}}_i$}. 
In our design, instead of trying to determine this assignment row by row, we make the assignment in ``blocks'' of rows. Namely,  each $\mathcal{W}_{i}$  will include  blocks of rows from $\boldsymbol{\tilde{X}}_i$.  Furthermore, we will use a common set of blocks for  the assignment of all $\mathcal{W}_{i}, i \in {[Z]}$,  which we refer to as row sets. These are  formally defined below.

\begin{defn}
Since each cs-matrix  has $\frac{q}{L}$ rows, we  partition the full index set of these rows $\{1,2, \cdots, \frac{q}{L} \}$ into $F$ consecutive disjoint subsets, possibly with varying sizes, called row sets, denoted by $\boldsymbol{\mathcal{M}}_t  = (\mathcal{M}_{1},\ldots,\mathcal{M}_{F} )$, whose union gives the full index set. 
\hfill $\Diamond$
 \end{defn}
 \begin{defn}
 Given row sets $\boldsymbol{\mathcal{M}}_t  = (\mathcal{M}_{1},\ldots,\mathcal{M}_{F} )$, we define  cs-matrix sets
  $\boldsymbol{\mathcal{P}}_t = (\mathcal{P}_{1} , \ldots , \mathcal{P}_{F} )$ where each $\mathcal{P}_{f}$ includes the indices  of $L$ cs-matrices for which all rows in  $\mathcal{M}_{f}$ are computed by the machines that store these cs-matrices. 
    Specifically, if $i \in \mathcal{P}_{f}$ and a machine $n$  stores   $\boldsymbol{\tilde{X}}_i$ ($i \in \mathcal{Q}_n$), then machine $n$ will compute  all rows in $\mathcal M_f$ from $\boldsymbol{\tilde{X}}_i$, i.e.,  these rows are included in $\mathcal{W}_{i}$.     This ensures that each row set $\mathcal{M}_{f}$  is computed  exactly $L$ times using the $L$  cs-matrices stored on these machines.
   \hfill $\Diamond$
 \end{defn}

From the above definitions, we see that  the rows computed by machine $n\in \mathcal{N}_t$ in time step $t$ are in the set
\be
\bigcup_{i\in\mathcal{Q}_n}\mathcal{W}_{i} = \bigcup_{i\in\mathcal{Q}_n}\left\{ \mathcal{M}_f : f\in[F],  i \in \mathcal{P}_{f}\right\}.
\ee
Note that the sets $\mathcal{M}_{1},\ldots,\mathcal{M}_{F}$ and $\mathcal{P}_{1} , \ldots , \mathcal{P}_{F}$ and $F$ may vary with each time step.
\begin{remark} 
When each machine only stores one cs-matrix, there is a one-to-one mapping between a cs-matrix and a machine, and thus $\mathcal P_f$ also represents the  set of machines that compute rows in $\mathcal M_f$. This is used in Example 1 and Algorithm 1, where we assume that the machines have heterogeneous speeds but homogeneous storage. 
\end{remark}
\begin{remark} 
 The row set and cs-matrix set pair $\left(\boldsymbol{\mathcal{M}}_t, \boldsymbol{\mathcal{P}}_t \right)$ combine to determine the computation assignment. The computation load  $\boldsymbol{\mu}$ is a function of  $\left(\boldsymbol{\mathcal{M}}_t, \boldsymbol{\mathcal{P}}_t \right)$.
\end{remark}

One example to illustrate $\Mc_f$ and $\Pc_f$ is given in Fig.~\ref{fig: exp1}(a). In this example, we consider the case that all machines have heterogeneous computing speed but with homogeneous storage, i.e., $\sigma[n]=1, n \in [N]$. Each machine store one cs-matrix and the union of $\Mc_f, f \in [4]$ in different colors cover all the row indices, $\left[ \frac{q}{L}\right]$. We let machine $n$ stores $\boldsymbol{\tilde{X}}_n$, then we can see that $\Pc_1 = \{1,5,6\}$, $\Pc_2 = \{2,3,4\}$, $\Pc_3 = \{3,5,6\}$, $\Pc_4 = \{4,5,6\}$. In this case, since there is a one-to-one mapping between the machine and the cs-matrix that it stores, the cs-matrix set $\mathcal P_f$ can also be interpreted as an index set of the machines that compute the stored cs-matrices. 

\subsection{Formulation of a Combinatorial Optimization Framework}
In a given time step $t$, our goal is to define the computation assignment, $\boldsymbol{\mathcal{M}}_t$ and $\boldsymbol{\mathcal{P}}_t $, such that the resulting computation load vector defined in (\ref{eq: compload_vector}) has the minimum computation time.
In time step $t$, given $\mathcal{N}_t$, $\mathcal{Q}_1,\ldots,\mathcal{Q}_N$ and $\boldsymbol{s}$, the optimal computation time, $c^*$, is the minimum of computation time defined by all possible computation assignments, $\left(\boldsymbol{\mathcal{M}}_t, \boldsymbol{\mathcal{P}}_t \right)$. 
Hence, 
based on all the conditions discussed before and given the storage vector $\boldsymbol{\sigma}$, 
we can formulate the following combinatorial optimization problem.
\begin{subequations} \label{eq: optprob_assign}
\begin{align}
\underset{{\left(\boldsymbol{\mathcal{M}}_t, \boldsymbol{\mathcal{P}}_t \right)} }{\text{minimize}} & \quad c\left(\boldsymbol{\mu}\left(\boldsymbol{\mathcal{M}}_t, \boldsymbol{\mathcal{P}}_t \right)\right) \label{eq: opt 1} \\
\text{subject to:} & \bigcup_{\mathcal{M}_f \in \boldsymbol{\mathcal{M}}_t}\mathcal{M}_f=\left[ \frac{q}{L}\right],\label{eq: optprob_assign 0}\\
& \mathcal{P}_f\subseteq\bigcup_{n\in\mathcal{N}_t}\mathcal{Q}_n,\;\; \forall \mathcal{P}_f \in \boldsymbol{\mathcal{P}}_t, \label{eq: optprob_assign 1}\\
& |\mathcal{P}_f|= L, \;\; \forall \mathcal{P}_f \in \boldsymbol{\mathcal{P}}_t,\label{eq: optprob_assign 2} \\
& |\boldsymbol{\mathcal{M}}_t|=|\boldsymbol{\mathcal{P}}_t| \label{eq: optprob_assign 3},
\end{align}
\end{subequations}

 The objective function (\ref{eq: opt 1}) is the overall computation time of a computation load vector $\boldsymbol{\mu}$, which is a function of $\left(\boldsymbol{\mathcal{M}}_t, \boldsymbol{\mathcal{P}}_t \right)$. Conditions (\ref{eq: optprob_assign 0})-(\ref{eq: optprob_assign 3}) specifies the constraints on $\left(\boldsymbol{\mathcal{M}}_t, \boldsymbol{\mathcal{P}}_t \right)$, which are to be optimized over. Specifically,  (\ref{eq: optprob_assign 0}) ensures that the union of the row sets $\mathcal M_f$ equals the full set of $\frac{q}{L}$ rows. i.e., all rows are assigned to be computed by some active machines at time $t$.
Condition (\ref{eq: optprob_assign 1}) ensures that the cs-matrices in $\mathcal P_f$  are stored by only  active machines at time $t$, and hence, each row set $\mathcal M_f$ is only assigned to be computed from these active machines. 
 Condition (\ref{eq: optprob_assign 2}) ensures that each row set is computed from exactly $L$ cs-matrices. Condition (\ref{eq: optprob_assign 3}) ensures that each row set has a corresponding cs-matrix set, i.e., the number of row sets equals the number of cs-matrix set.

\begin{remark}
In Sections~\ref{sec: compload} and \ref{sec: compassign}, we precisely solve the combinatorial optimization problem of (\ref{eq: optprob_assign}). 
We decompose this problem into two sub-problems. First, a convex optimization problem to find an optimal computation load vector $\boldsymbol{\mu}$ without the consideration of a specific computation assignment. This means that we will solve the problem of (\ref{eq: optprob_assign}) by treating $\boldsymbol{\mu}$ as a real vector without considering whether such a computation assignment is feasible. Second, given the optimal $\boldsymbol{\mu}$ solved in the previous convex optimization problem, we solve a computation assignment problem or a ``filling problem" in order to find a $\left(\boldsymbol{\mathcal{M}}_t, \boldsymbol{\mathcal{P}}_t \right)$ that meets the optimal computation load. Moreover,  we show that an optimal assignment, $\left(\boldsymbol{\mathcal{M}}_t, \boldsymbol{\mathcal{P}}_t \right)$, can be found via a low complexity algorithm that completes in at most $N_t$ iterations and the number of computation assignments, $F$, is at most $N_t$.
\end{remark}

Before going into the details of the solution for the optimization problem  (\ref{eq: optprob_assign}), we will first present two toy examples to illustrate the proposed algorithms.

\section{Two CEC Examples}
\label{sec: example}

In this section, we will discuss two CEC examples to illustrate the proposed approaches. Example 1  considers the scenario where machines have heterogeneous computing speeds but  homogeneous storage constraints. Example 2 considers the more general scenario where machines have both heterogeneous storage space and computing speeds.

\subsection{Example 1: 
CEC with Heterogeneous computing speeds/Homogeneous Storage Constraints}
\label{sec: Machines With Homogeneous Storage Constraints}

We consider a system with a total of $N=6$ machines where each has the storage capacity to store $\frac{1}{3}$ of the data matrix $\boldsymbol{X}$. In time step $t$, the machines have the collective goal of computing $\boldsymbol{y}_t=\boldsymbol{X}\boldsymbol{w}_t$ where $\boldsymbol{w}_t$ is some vector. In order to allow for preempted machines, $\boldsymbol{X}$ is split row-wise into $L=3$ sub-matrices, $\boldsymbol{X}_1$, $\boldsymbol{X}_2$ and $\boldsymbol{X}_3$ and an MDS code is used to construct the cs-matrices $\{\boldsymbol{\tilde{X}}_n : n\in[N] \}$ which are stored among the machines. In particular, when the storage space among machines is homogeneous, machine $n$ stores only one cs-matrix $\boldsymbol{\tilde{X}}_n$ and $\boldsymbol{\sigma}=[1,\ldots,1]$.
In this case, there is a one-to-one mapping between cs-matrices and machines. 
This placement is designed such that any element of $\boldsymbol{y}_t$ can be recovered by obtaining the corresponding coded computation from any $L=3$ machines. 
To recover the entirety of $\boldsymbol{y}_t$, we split the  cs-matrices into row sets, such that each set is used for computation at $L=3$ machines.


The machines have relative computation speeds defined by 
$\boldsymbol{s} = [\;2,\;\;2,\;\;3,\;\;3,\;\;4,\;\;4\;]$, 
where  machines $5$ and $6$ are the fastest  and can perform row computations twice as fast as machines $1$ and $2$. Machines $3$ and $4$ are the next fastest  and can perform matrix computations $1.5$ times as fast as machines $1$ and $2$. Our goal is to assign computations, or rows of the  cs-matrices, to the machines to minimize the overall computation time with the constraint that each computation is assigned to $3$ machines.

\begin{figure*}
\centering \includegraphics[width=15cm]{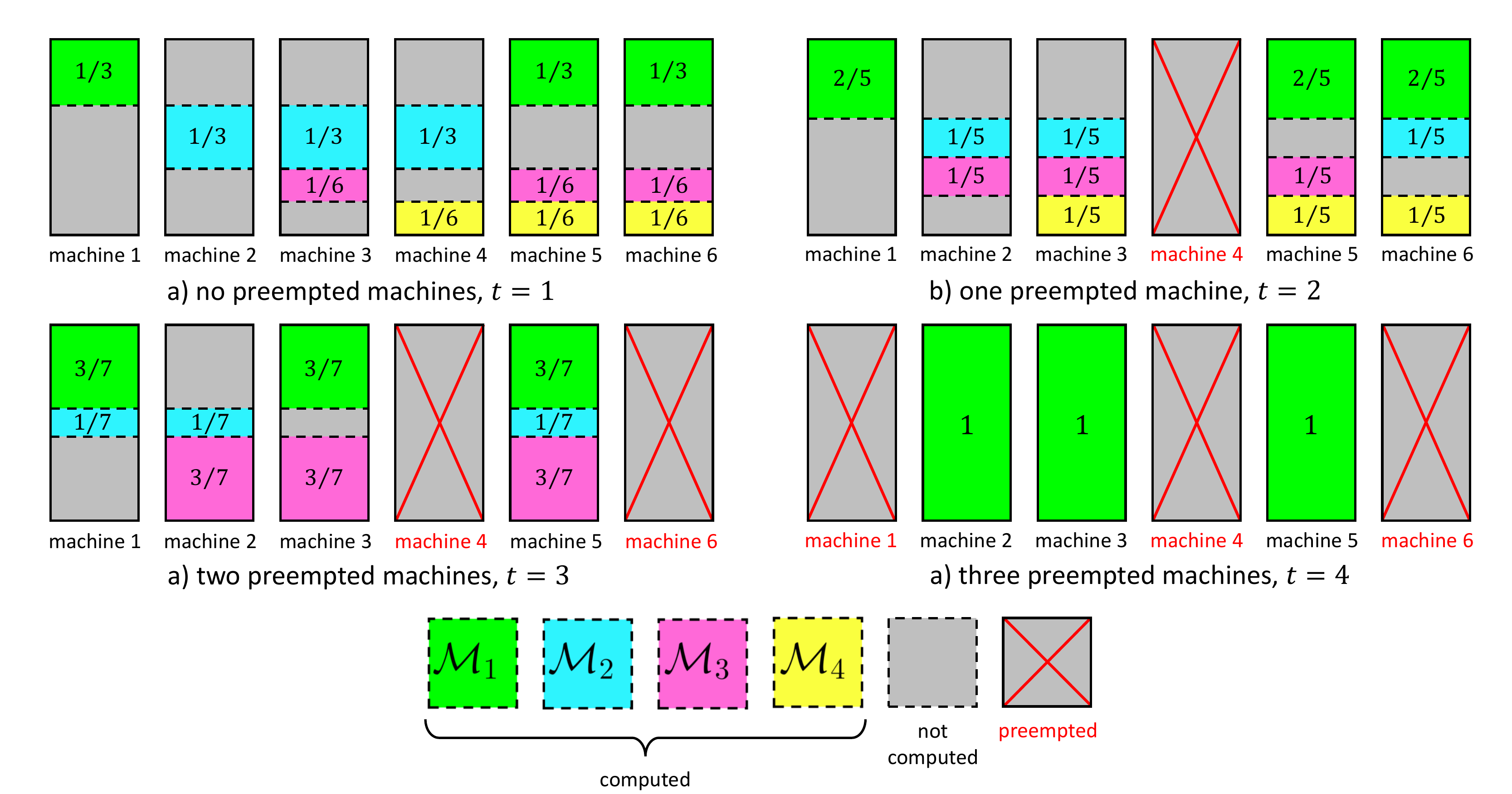} 
\vspace{-0.6cm}
\caption{~\small An illustration of the optimal computation assignments in Example 1 over $4$ times steps on a heterogeneous CEC network where machines have heterogeneous computing speed and homogeneous storage space. In this example, $N=6$, $Z=6$, $L=3$ and $\Qc_n=\{ n \}, \forall n \in [4]$.   At $t=1$, there are $F=4$ row sets $\mathcal M_1$ (green)  $\mathcal M_2$ (blue),  $\mathcal M_3$ (magenta),  $\mathcal M_4$ (yellow), each is assigned to $L=3$ cs-matrices. The number labeled  in the center of each row set is the fraction of the rows in that row set. The row sets change over time. At $t=3$, there are $F=3$ row sets, and at $t=4$, there is only  $F=1$ row set. }
\label{fig: exp1}
\end{figure*}

In time step $1$, there are no preempted machines which means that $\mathcal{N}_1=\{1,\ldots, 6 \}$ and $N_1=6$. We assign fractions of the rows to the machines defined by the computation load vector 
$\boldsymbol{\mu} = \left[\;\frac{1}{3},\;\;\frac{1}{3},\;\;\frac{1}{2},\;\;\frac{1}{2},\;\;\frac{2}{3},\;\;\frac{2}{3}\;\right]$, 
such that machines $1$ and $2$ are assigned $\frac{1}{3}$, machines $3$ and $4$ are assigned $\frac{1}{2}$ and machines $5$ and $6$ are assigned $\frac{2}{3}$ of the rows of their respective cs-matrices. We define $\boldsymbol{\mu}$ such that it sums to $L=3$ and each row can be assigned to $3$ machines. 
Furthermore, based on the machine computation speeds, the machines finish at the same time to minimize the overall computation time. 
In Section \ref{sec: compload}, we outline 
the systematic approach to determine $\boldsymbol{\mu}$. Next, given $\boldsymbol{\mu}$, the rows of the  cs-matrices must be assigned. We define row sets, $\mathcal{M}_{1}$, $\mathcal{M}_{2}$, $\mathcal{M}_{3}$, and $\mathcal{M}_{4}$ which are assigned to $F=4$ sets of $L=3$ cs-matrices  $\mathcal{P}_{1}$, $\mathcal{P}_{2}$, $\mathcal{P}_{3}$, and $\mathcal{P}_{4}$. Since each machine $n$ stores one  cs-matrix $\boldsymbol{\tilde{X}}_n$, it is equivalent to say that we assign computations to machines. In other words $\mathcal{P}_{1}$, $\mathcal{P}_{2}$, $\mathcal{P}_{3}$, and $\mathcal{P}_{4}$ represent the machines assigned to compute $\mathcal{M}_{1}$, $\mathcal{M}_{2}$, $\mathcal{M}_{3}$, and $\mathcal{M}_{4}$, respectively. These sets are depicted in Fig.~\ref{fig: exp1}(a) where, for example, $\mathcal{M}_{1}$ contains the first $\frac{1}{3}$ of the rows assigned to machines $\mathcal{P}_{1} = \{ 1,5,6\}$. Moreover, $\mathcal{M}_{2}$ contains the next $\frac{1}{3}$ of the rows assigned to machines $\mathcal{P}_{2}=\{2,3,4\}$, $\mathcal{M}_{3}$ contains the next $\frac{1}{6}$ of the rows assigned to machines $\mathcal{P}_{3}=\{3,5,6\}$ and $\mathcal{M}_{4}$ contains the final $\frac{1}{6}$ of the rows assigned to machines $\mathcal{P}_{4}=\{4,5,6\}$. In Section \ref{sec: compassign}, we present Algorithm \ref{algorithm:1}  to determine the computation assignment for general $\boldsymbol{\mu}$. By this assignment, the fraction of rows assigned to machine $n$ sums to $\mu[n]$ and each row is assigned to $L=3$ machines to recover the entirety of $\boldsymbol{y}_1$.

In time step $2$, we find $N_2=5$ because machine $4$  is preempted and
no longer available to perform computations. Therefore, the computations must be re-assigned among $\mathcal{N}_2=\{1, 2, 3, 5, 6 \}$. First, we obtain 
$\boldsymbol{\mu} = \left[\;\frac{2}{5},\;\;\frac{2}{5},\;\;\frac{3}{5},\;\;0,\;\;\frac{4}{5},\;\;\frac{4}{5}\;\right]$, 
which sums to $L=3$ and minimizes the overall computation time. Given $\boldsymbol{\mu}$, we then use Algorithm \ref{algorithm:1} (see Section~\ref{sec: compassign}) to assign computations to a machine with the least number of remaining rows to be assigned and $L-1=2$ machines with the most number of remaining rows to be assigned. For example, in the first iteration, $\mathcal{M}_{1}$ is defined to contain the first $\frac{2}{5}$ of the rows and is assigned to machines $\mathcal{P}_{1}= \{ 1,5,6\}$. After this iteration, machines $2$, $5$ and $6$ require $\frac{2}{5}$ of the total rows to still be assigned to them and machine $3$ requires $\frac{3}{5}$ of the total rows. In the next iteration, $\mathcal{M}_{2}$ contains the next $\frac{1}{5}$ of the rows and is assigned to $\mathcal{P}_{2}=\{ 2,3,6\}$. Note that, only $\frac{1}{5}$ of the rows could be assigned in this iteration otherwise there would only be two machines, $3$ and $5$, which still require assignments and therefore, the remaining rows cannot be assigned to three machines. In the final two iterations, $\mathcal{M}_{3}$ and $\mathcal{M}_{4}$ contain $\frac{1}{5}$ of the previously unassigned rows and are assigned to the machines of $\mathcal{P}_{3} = \{2,3,5 \}$ and $\mathcal{P}_{4} = \{3,5,6 \}$, respectively. These assignments are depicted in Fig.~\ref{fig: exp1}(b).

Next, in time step $3$, we find $N_3=4$ because machines $4$ and $6$ are preempted. Hence, $\mathcal{N}_3=\{1, 2, 3, 5 \}$. Similar to previous time steps, it is ideal to have machines $3$ and $5$ compute $1.5\times$ and $2\times$ the number of computations, respectively, compared to machines $1$ and $2$. However, this is not possible since each machine can be assigned at most a number of rows equal to the number of rows of the cs-matrices. In this case, we assign all rows to the fastest machine, machine $5$, and assign fractions of the rows to the remaining machines which sum up to $2$. As a result, we let 
$\boldsymbol{\mu} = \left[\;\frac{4}{7},\;\;\frac{4}{7},\;\;\frac{6}{7},\;\;0,\;\;1,\;\;0\;\right]$. 
Then, Algorithm \ref{algorithm:1} defines $\mathcal{M}_{1}$, $\mathcal{M}_{2}$ and $\mathcal{M}_{3}$ as disjoint sets containing $\frac{3}{7}$, $\frac{1}{7}$ and $\frac{3}{7}$ of the rows, respectively. Moreover, these row sets are assigned to the machines of $\mathcal{P}_{1}= \{ 1,3,5\}$, $\mathcal{P}_{2}= \{ 1,2,5\}$ and $\mathcal{P}_{3}= \{ 2,3,5\}$, respectively. These assignments are depicted in Fig.~\ref{fig: exp1}(c).

Finally, in time step $4$, machines $1$, $4$ and $6$ are preempted which means that $\Nc_4 = \{2,3,5\}$ and $N_4=3$. To assign all the rows to $L=3$ machines, each available machine is assigned by all of the rows and 
$\boldsymbol{\mu} = \left[\;0,\;\;1,\;\;1,\;\;0,\;\;1,\;\;0\;\right]$.
In other words, $\mathcal{M}_{1}$ contains all rows and $\mathcal{P}_{1}=\{2,3,5 \}$. This is depicted in Fig.~\ref{fig: exp1}(d).

Next, we  present  Example 2  with heterogeneous storage space and computing speeds. This example uses Algorithm 2,  which is a generalization of  Algorithm 1  discussed in Example 1.

\subsection{Example 2: CEC with Heterogeneous computing speeds and Storage Constraints}
\label{sec: Machines With Heterogeneous Storage Constraints}

\begin{figure*}
\centering
\centering \includegraphics[width=16cm]{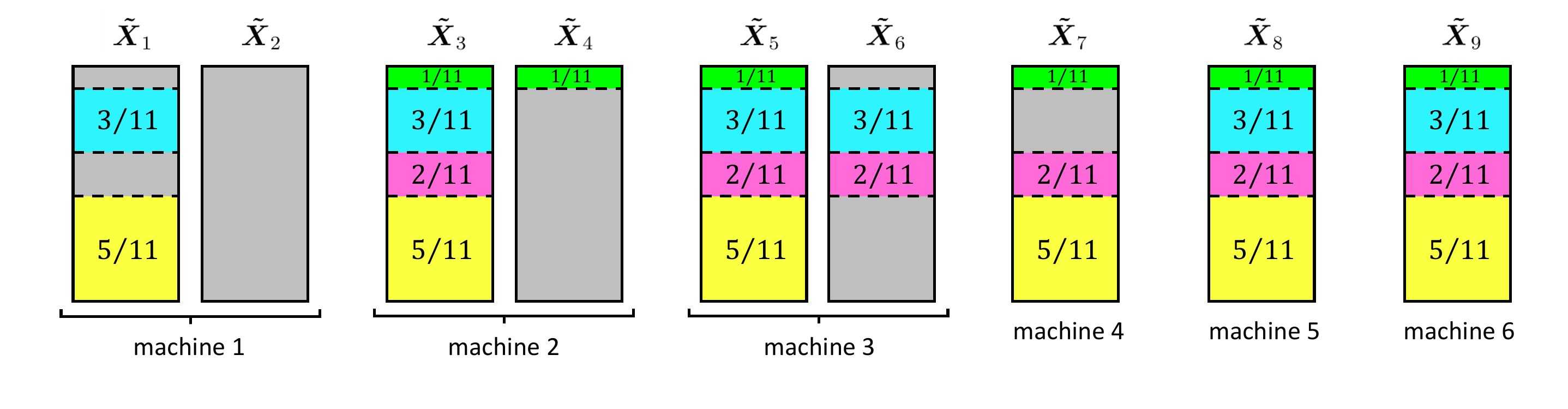} 
\vspace{-0.8cm}
\caption{~\small Example 2: Optimal computation assignments for a CEC network with machines that have heterogeneous storage requirements and varying computations speeds.   Here, we have $N=6$, $Z=9$, $L=6$. The machines have varying computation speeds $\boldsymbol{s} = [2,\; 3, \; 4, \; 2, \; 3, \; 4\;]$ and storage capacity
$\boldsymbol{\sigma} = [2,\; 2, \; 2, \; 1, \; 1, \; 1 \;].$ Machines 1 to 3 each stores 2 cs-matrices, and machines 4 to 6 each stores only 1 cs-matrix. The machines are ordered in the decreasing order of SCR $\frac{\sigma[n]}{s[n]}$. There are 4 row sets $\mathcal M_1$ (green)  $\mathcal M_2$ (blue),  $\mathcal M_3$ (magenta),  $\mathcal M_4$ (yellow), each is assigned to $L=6$ cs-matrices. For instance, $\mathcal M_1$ is assigned to cs-matrices $\mathcal P_1=\{3,4,5,7,8,9\}$. The optimal computation load vector is $\boldsymbol{\mu} = [\;\frac{8}{11},\;\;\frac{12}{11},\;\;\frac{16}{11},\;\;\frac{8}{11},\;\;1,\;\;1\;]$.  For instance, machine 3 computes one full cs-matrix $\boldsymbol{\tilde{X}}_5$ and partially computes $\frac{3}{11} +\frac{2}{11}=\frac{5}{11}$ of $\boldsymbol{\tilde{X}}_6$, which adds up to a computation load of $\mu[3]=1+\frac{5}{11}=\frac{16}{11}$.}
\label{fig: exp2}
\vspace{-0.5cm}
\end{figure*}

Consider the case where $L=6$ and there are $N=6$  machines which each have distinct speed-storage pairs. For ease of presentation, we only focus on a single time step (or $t=1$) and assume there are no preempted machines ($N_1=6$). 
The 
computing speed of the available machines are defined by
$\boldsymbol{s} = \left[\;2,\;\; 3, \;\; 4, \;\; 2, \;\; 3, \;\; 4\; \right]$.
The total number of stored  cs-matrices is $Z_1=9$ and the machines store a number of  cs-matrices defined by 
$\boldsymbol{\sigma} = \left[\;2,\;\; 2, \;\; 2, \;\; 1, \;\; 1, \;\; 1 \;\right]$.
For example, machine 1 has a speed of 2 and stores 2  cs-matrices and machine 5 has a speed of 3 and stores 1  cs-matrix. 
The storage of the cs-matrices at each machine is shown in Fig.~\ref{fig: exp2} where $\boldsymbol{\tilde{X}}_i$ is labeled at the top of each block which represent a cs-matrix. 
The machines are in descending order based on $\frac{\sigma[n]}{s[n]}$ to use Theorem 1 to find the optimal computation load vector described in Section \ref{sec: compload}. Based on Theorem 1, the computation load vector is 
$\boldsymbol{\mu} = \left[\;\frac{8}{11},\;\;\frac{12}{11},\;\;\frac{16}{11},\;\;\frac{8}{11},\;\;1,\;\;1\;\right]$.
Notice that, different from  Example 1,  a machine here may have a computation load greater than $1$ if it performs computations on more than $1$ stored cs-matrices. However, similar to Example 1, machines either have the same computation time or perform computations on all locally stored data. In this example, based on the computation speeds, machines $1$ through $4$ complete the assigned computing tasks at the same time and machines $5$ and $6$ compute using the entirety of their one locally available cs-matrix and finish before the other machines.

Next, we need to assign computations that yield the computation load vector, $\boldsymbol{\mu}$. 
We use Algorithm 2 in Section~\ref{sec: compassign}, where
instead of assigning computations to machines, we assign computations to each cs-matrix at each machine. 
In this algorithm, we need to first decide how much of each cs-matrix will be computed. For example, consider machine $3$ which has a computation load of $\mu[3]=\frac{16}{11}$ and locally stores $\sigma[3]=2$ cs-matrices. There is a choice of how much machine $3$ will compute each of its  cs-matrices. A solution which simplifies the assignment is for machine $3$ to compute the entirety of one cs-matrix and a $\frac{5}{11}$ fraction of the other. Similarly, machine $2$ will compute the entirety of one cs-matrix and $\frac{1}{11}$ fraction of the other. In general, when $\sigma[n]>1$ and machine $n$ stores more than one cs-matrices, it will compute $\lfloor \mu[n] \rfloor$ whole cs-matrices and a $\mu[n]-\lfloor \mu[n] \rfloor$ fraction of the remaining cs-matrix.

The final computation assignments are shown in Fig. \ref{fig: exp2}. There are $F=4$ matrix row sets $\mathcal{M}_1$ through $\mathcal{M}_4$ which contain a $\frac{1}{11}$, $\frac{3}{11}$, $\frac{2}{11}$ and $\frac{5}{11}$ fraction of rows, respectively. Each row set $\mathcal M_f$ is assigned to a sc-matrix set $\mathcal P_f$ that contains $L=6$ sc-matrices. These are given by sc-matrix sets $\mathcal P_1=\{ 3,4,5,7,8,9\}$, $\mathcal P_2=\{ 1,3,5,6,8,9\}$, $\mathcal P_3=\{ 3,5,6,7,8,9\}$, and $\mathcal P_4=\{ 1,3,5,7,8,9\}$.  

\section{First sub-problem: Optimal Computation Load Vector}
\label{sec: compload}

We decompose the optimization problem (\ref{eq: optprob_assign}) into two sub-problems. 
In this section, we present the first sub-problem by
introducing a relaxed convex optimization problem 
to find the optimal computation load vector $\boldsymbol{\mu^*}$ and its corresponding computation time ${\hat c}{^*}=c(\boldsymbol \mu^*)$ 
without considering an explicit computation assignment $(\boldsymbol{\mathcal{M}}_t,\boldsymbol{\mathcal{P}}_t)$. Due to the relaxed constraints, we have $\hat{c}^*\leq c^*$. 
Next, in Section~\ref{sec: compassign}, we will present the second sub-problem in which 
we show that it is possible to find a computation assignment $(\boldsymbol{\mathcal{M}}_t,\boldsymbol{\mathcal{P}}_t)$ that achieves the $\boldsymbol{\mu^*}$ that we found in the first step. Hence, there is no gap between the ``relaxed" convex optimization problem and (\ref{eq: optprob_assign}) and we have $c^* = \hat{c}^*$.
\subsection{The Proposed Relaxed Convex Optimization Problem}

Given a computation speed vector $\boldsymbol{s}$ and storage vector $\boldsymbol{\sigma}$, we let the optimal computation load vector $\boldsymbol{\mu}^*$ be the solution to the following convex 
optimization problem:
\begin{subequations} \label{eq: optprob}
\begin{align}
\underset{\boldsymbol{\mu}}{\text{minimize }} \; & c(\boldsymbol{\mu})= \max_{n\in [N_t]}\frac{\mu[n]}{s[n]} \\
\text{subject to:} &\sum_{n\in [N_t]}\mu[n] = L, 
\label{eq:sum_L}\\
&0 \leq \mu[n] \leq \sigma[n], \forall n \in [N_t], \\
& \mu[n] \in \mathbb{R}^+, \; \forall n\in [N_t]
\end{align}
\end{subequations}
which can be shown to be a convex optimization problem.
While computation assignments, $(\boldsymbol{\mathcal{M}}_t,\boldsymbol{\mathcal{P}}_t)$, are not explicitly considered in (\ref{eq: optprob}), we note that the key constraint of $\sum_{n\in [N_t]}\mu[n] = L$ is a relaxed version of the requirement on the computation assignment that each row set should be assigned to $L$ sc-matrices. 
It is important to note that the analytical solution to the optimization problem (\ref{eq: optprob}) can be explicitly found.
When $Z_t=L$, it can be seen that this optimal solution is given by $\boldsymbol{\mu}^* = \boldsymbol{\sigma}$. 
When $Z_t>L$, the  analytical optimal solution to (\ref{eq: optprob}) is presented in the following theorem.
\begin{theorem}\label{th: load_assignment}
Assume that $Z_t>L$ and { the machines are labeled in the decreasing order of the storage capacity to computation speed ratio (SCR)} 
\be
\label{eq: theorem 1 1}
\frac{\sigma[1]}{s[1]} \geq \frac{\sigma[2]}{s[2]} \geq \cdots \geq\frac{\sigma[N_t]}{s[N_t]}.
\ee
    The optimal solution $\boldsymbol \mu^*$ to  the optimization problem of (\ref{eq: optprob}) must take the following form
\begin{equation}
\mu^*[n]=\begin{cases}
\hat c^* s[n] & \text{if } 1\le n \le k^*\\
\sigma[n] & \text{if } k^*+1 \le n \le N_t,
\end{cases}
\label{eq:optimal_form}
\end{equation}
where $k^*$ is the largest integer in $[N_t]$ such that
\be
\frac{\sigma[k^*+1]}{s[k^*+1]} < \hat c^* = \frac{L-\sum_{n=k^*+1}^{N_t}\sigma[n]}{\sum_{n=1}^{k^*}s[n]} \leq \frac{\sigma[k^*]}{s[k^*]},\quad  \text{if $k^* < N_t$},
\label{eq: c_bounds-fork}
\ee
otherwise, $k^* = N_t$.
Here, ${\hat c}{^*}=c(\boldsymbol \mu^*)$ is the maximum computation time among the $N_t$ machines given the computation load assignment  $\boldsymbol \mu^*$.
\end{theorem}

\subsection{Proof of Theorem~\ref{th: load_assignment}} 
In the following, we first present two Claims that will lead to the proof of Theorem~\ref{th: load_assignment}. 

\begin{claim}\label{cl: 1}
{If $\mu^*[n]<\hat{c}^* s[n]$, then $\mu^*[n] = \sigma[n]$. Thus, in this case the optimal computation load assigned to machine $n$ is equivalent to its storage.}

\end{claim}

\begin{IEEEproof}
We prove Claim \ref{cl: 1} by contradiction.
Since $\hat c^*=\text{max}_{n \in [N_t]}\frac{\mu^*[n]}{s[n]}$, we define two disjoint sets $\mathcal{T}_0$ and $\mathcal{T}_1$, where $\mathcal{T}_0 \bigcup \mathcal{T}_1=[N_t]$, as follows.
\be
\mathcal{T}_0=\{n\in[N_t]:\mu^*[n]=\hat{c}^*  s[n]\}
\ee
and
\be
\mathcal{T}_1=\{n\in[N_t]:\mu^*[n]<\hat{c}^*  s[n]\}.
\ee
In the following, we will show that if there exists an $i \notin \mathcal{T}_0$, then we must have $i\in \mathcal{T}_1$ and $\mu^*[i]=\sigma[i]$. In order to do this, we will construct a new solution $\boldsymbol{\mu}'$ from the optimal solution $\boldsymbol{\mu}^*$ such that $c(\boldsymbol{\mu}') < \hat{c}^*$, which leads to a contradiction that  $\boldsymbol{\mu}^*$ is an optimal solution. The details are as follows. Assume that there exists some $i\in[N_t]$ such that $ i \in \mathcal{T}_1$ and $\mu^*[i] < \sigma[i]$.
Define $\boldsymbol{\mu}'$ such that
\begin{align}
\mu'[n] = \left\{ \begin{array}{cc}
                \mu^*[n] + \epsilon  \hspace{5mm} &\text{if }n=i, \\
                \mu^*[n] - \frac{\epsilon}{|\mathcal{T}_0|}  \hspace{5mm} &\text{if }  n\in\mathcal{T}_0, \\
                \mu^*[n]  \hspace{5mm} &\text{if } n \in  \mathcal{T}_1 \setminus i\\
                \end{array} \right.
\end{align}
where $0<\epsilon<\sigma[n]-\mu^*[n]$ and $\epsilon$ is sufficiently small such that
\be
\frac{\mu'[i]}{s[i]} = \frac{\mu^*[i]+\epsilon}{s[i]} < \hat{c}^*,
\ee
and for all $n \in \mathcal{T}_0$
\be
 \mu^*[n] - \frac{\epsilon}{|\mathcal{T}_0|} > 0.
\ee
One can verify that we have $\frac{\mu'[n]}{s[n]} < \hat{c}^*$ for any $n \in [N_t]$ and thus we obtain $c(\boldsymbol{\mu}') < \hat{c}^*$. This contradicts with the assumption that $\boldsymbol{\mu}^*$ is optimal. Thus, it follows that if $n \notin \mathcal{T}_0$, then we must have $n\in \mathcal{T}_1$ and $\mu^*[n]=\sigma[n]$.
\end{IEEEproof}

\begin{claim}\label{cl: 2}
  If $j \in \mathcal{T}_0 $ and $i \in \mathcal{T}_1$, then
   $\frac{\sigma[j]}{s[j]} > \frac{\sigma[i]}{s[i]}$.
\end{claim}
\begin{IEEEproof}
This claim  follows directly from
\be\label{eq: pfeq1}
\frac{\mu^*[i]}{s[i]}=\frac{\sigma[i]}{s[i]} < \hat{c}^*= \frac{\mu^*[j]}{s[j]} \leq \frac{\sigma[j]}{s[j]}.
\ee
\end{IEEEproof}

\indent {\it{Proof of  Theorem~\ref{th: load_assignment}:}}
Combining Claims \ref{cl: 1} and \ref{cl: 2}, we find that the optimal solution must take the form of
\begin{equation}
\mu^*[n]=\begin{cases}
\hat c_k^* s[n] & \text{if } 1\le n \le k,\\
\sigma[n] & \text{if } k+1 \le n \le N_t,
\end{cases}
\label{eq:optimal_form_k}
\end{equation}
where $k=|\mathcal{T}_0|$. Next, we will optimize $k$ such that $\hat c_k^*$ is minimized.
Since $\sum_{n\in [N_t]}\mu^*[n] = L$, by using (\ref{eq:optimal_form_k}), we obtain (\ref{eq: c_bounds-fork}) because
\begin{align}
L = \sum_{n=1}^{N_t}\mu^*[n]
  &=  \sum_{n=1}^{k}\mu^*[n] + \sum_{n=k+1}^{N_t}\sigma[n] \\
  &=  \hat{c}_k^*\sum_{n=1}^{k}s[n]  + \sum_{n=k+1}^{N_t}\sigma[n]
\end{align}
and
\be \label{eq: c_k}
\hat{c}_k^* = \frac{L - \sum_{n=k+1}^{N_t}\sigma[n]}{\sum_{n=1}^{k}s[n]}.
\ee
The left-most inequality of (\ref{eq: c_bounds-fork}) follows from  $k \in \mathcal{T}_0$ and $\mu^*[k] \le \sigma[n]$. The right-most inequality  of (\ref{eq: c_bounds-fork}) follows from  $k+1 \in \mathcal{T}_1$ and $\mu^*[k+1] = \sigma[n]$.
 Since $\frac{\sigma[n]}{s[n]}$ is a decreasing sequence, we see from  (\ref{eq: c_bounds-fork}) that $\hat c_k^*$ is maximized when $k$ is chosen to be $k^*$,  the largest value in $[N_t]$ such that (\ref{eq: c_bounds-fork}) is satisfied.

\subsection{Discussions on Theorem~\ref{th: load_assignment}}
From (\ref{eq: theorem 1 1}), (\ref{eq:optimal_form}) and (\ref{eq: c_bounds-fork}), we can observe that the optimal solution $\boldsymbol \mu^*$ to the optimization problem (\ref{eq: optprob}) is always rational due to the fact that $\boldsymbol s$ are rational numbers and $\boldsymbol \sigma$ are integers. Hence, it is achievable for large enough $q$ if the computation assignment exists.
The following corollary presents the solution of the optimal computation load vector when the storage among machines is homogeneous, i.e., each machine stores exactly one cs-matrix. This storage design is equivalent to that used in the original CEC work of \cite{yang2018coded}, but here the machines have varying speeds  as opposed to the homogeneous setting of \cite{yang2018coded}.

\begin{corollary}\label{crly: 1}
When $\boldsymbol{\sigma}=[1,1\cdots,1]$, we find
\begin{equation}
\mu^*[n]=\begin{cases}
\hat c^* s[n] & \text{if } 1\le n \le k^*\\
1 & \text{if } k^*+1 \le n \le N_t,
\end{cases}
\label{eq:optimal_form_hm}
\end{equation}
where $k^*$ is the largest integer in $[N_t]$ such that
\be
\frac{1}{s[k^*+1]} < \hat c^* = \frac{L-N_t+k^*}{\sum_{n=1}^{k^*}s[n]} \leq \frac{1}{s[k^*]}.
\label{eq: c_bounds-fork_hm}
\ee
\end{corollary}
\begin{IEEEproof}
  Corollary \ref{crly: 1} is proved by substituting $\sigma[n]=1$ for $n\in[N_t]$ in equations (\ref{eq:optimal_form}) and (\ref{eq: c_bounds-fork}) and ordering the machines by speed in ascending order.
\end{IEEEproof}

\begin{remark}
The two cases in (\ref{eq:optimal_form}) are determined by whether a machine $n$ satisfies $ \mu^*[n]=\hat c^* s[n]$ or $ \mu^*[n]< \hat c^* s[n]$. For the  first case when $1 \le n \le k^*$, the equality is achieved and we must have $0< \mu^*[n]\le \sigma[n]$. {Among these $k^*$ machines, the computation load $ \mu^*[n]$ is proportional to the computation speed $s[n]$.} For the second case when $k^*+1 \le n \le N$, we have the strict  inequality and $ \mu^*[n]=\sigma[n]$. {The computation load $\mu^*[n]$ equals (thus is limited by) the storage $\sigma[n]$.} 
 The equality in (\ref{eq: c_bounds-fork}) ensures that $\sum_{n=1}^{N_t} \mu^*[n]=L$; the right-most inequality ensures that $\mu^*[n] \le \mu^*[k^*]=\hat c^* s[k^*] \le \sigma[n],$ for any $1\le n \le k^*$;  the left-most inequality ensures that for any $k^*+1 \le n \le N$, we have
 $ \mu^*[n]< \hat c^* s[n]$. Hence, the computation time $\hat c^*$ is equal to the local computation time of any of the $k^*$ machines with the largest SCR.
\end{remark}

Since the optimization problem of (\ref{eq: optprob}) aims to minimize a convex function on a closed and convex set, the existence of an optimal solution is guaranteed. This ensures the existence of
  some $k^* \in [N_t]$ such that (\ref{eq: c_bounds-fork}) is satisfied. In the following, we provide a numerical procedure to find $k^*$.  First, it is straightforward to verify that if the right-hand-side (RHS) inequality ``$\le $'' of (\ref{eq: c_bounds-fork}) is violated for $k^*=i$, then the  left-hand-side (LHS) inequality ``$<$'' of (\ref{eq: c_bounds-fork}) must hold for $k^*=i-1$. In other words, for any $i\in[N_t]$,
\be
\text{if } \hat c_i^* > \frac{\sigma[i]}{s[i]}, \text{then } \frac{\sigma[i]}{s[i]} < \hat c_{i-1}^*.
\label{eq:equivalent}
\ee
 where $\hat c_i^*$ (and $\hat c_{i-1}^*$) are defined by (\ref{eq: c_k}) for different values of $k^*$.
We first check $k^*=N_t$. If the RHS of (\ref{eq: c_bounds-fork}) holds, then we have $k^*=N_t$. Otherwise, it follows from (\ref{eq:equivalent}) that the LHS of (\ref{eq: c_bounds-fork}) must hold for $k^*=N_t-1$. If the RHS of (\ref{eq: c_bounds-fork}) also hold for $k^*=N_t-1$, then we have $k^*=N_t-1$. Otherwise, it follows from (\ref{eq:equivalent}) that the LHS of (\ref{eq: c_bounds-fork}) must hold for $k^*=N_t-2$. We continue this process by decreasing $k^*$ until we find one value of $k^*$ for which both sides of (\ref{eq: c_bounds-fork}) hold. This process is guaranteed to terminate before reaching $k^*=1$ for which the RHS of (\ref{eq: c_bounds-fork}) always hold. Hence, this establishes the procedure to find $k^*$ directly using (\ref{eq: c_bounds-fork}).

\begin{figure*}
\centering
\centering \includegraphics[width=16cm]{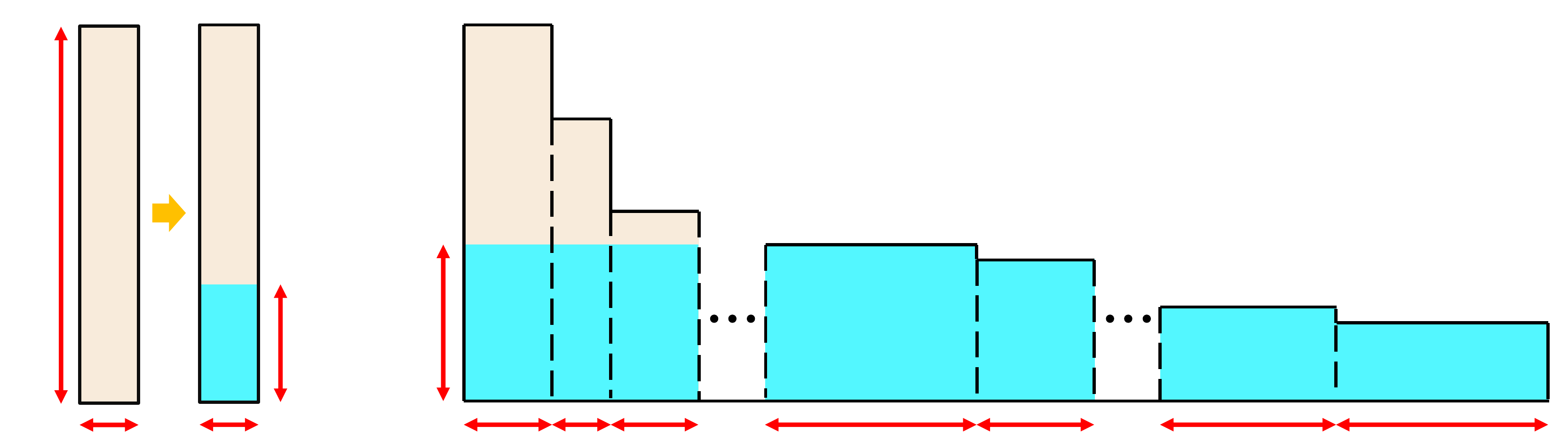} 
\put(-162,22){{\large $\frac{\sigma[n]}{s[n]}$}}
\put(-152,-4){{\small $s[n]$}}
\put(-140,-4){{\small $s[n]$}}
\put(-145,-10){{\small (a)}}
\put(-68,-10){{\small (b)}}
\put(-151,12){\rotatebox{90}{Area$=\sigma[n]$}}
\put(-130.5,9){{\large $\frac{\mu[n]}{s[n]}$}}
\put(-138.5,7){\rotatebox{90}{\small $\mu[n]$}}
\put(-119,10.5){$\hat{c}^*$}
\put(-111,-4){{\small $s[1]$}}
\put(-103.5,-4){{\small $s[2]$}}
\put(-96.5,-4){{\small $s[3]$}}
\put(-74,-4){{\small $s[k^*]$}}
\put(-78.5,22){{\small machine $k^*$}}
\put(-38,24){{\small Total fill area}}
\put(-43,20){{\small (across all machines)}}
\put(-12,22){{ $=L$}}
\put(-61,-4){{\small $s[k^*+1]$}}
\put(-39,-4){{\small $s[N_t-1]$}}
\put(-17,-4){{\small $s[N_t]$}}

\vspace{-0.cm}
\caption{~\small A water-filling like representation of the storage and computation load of (a) for machine $n$ only and (b) for a set of available machines with an optimal computation load vector that solves the optimization of (\ref{eq: optprob}) with the solution of Theorem \ref{th: load_assignment}.
Machines in (b) are ordered in the decreasing order of $\frac{\sigma(n)}{s(n)}$.
The storage of machine $n$ is represented by the area of the full rectangle (in peach color). The computation load $\mu[n]$ assigned to machine $n$ is represented by the area of the filled blue region  within the $n$-th rectangle. The height of the each filled blue region  represents $\frac{\mu[n]}{s[n]}$, which is the computation time of machine $n$. 
The maximum height of these filled regions represent the computation time $\hat c^*=c(\boldsymbol{\mu^*})$.
}
\label{fig: waterfill}
\end{figure*}
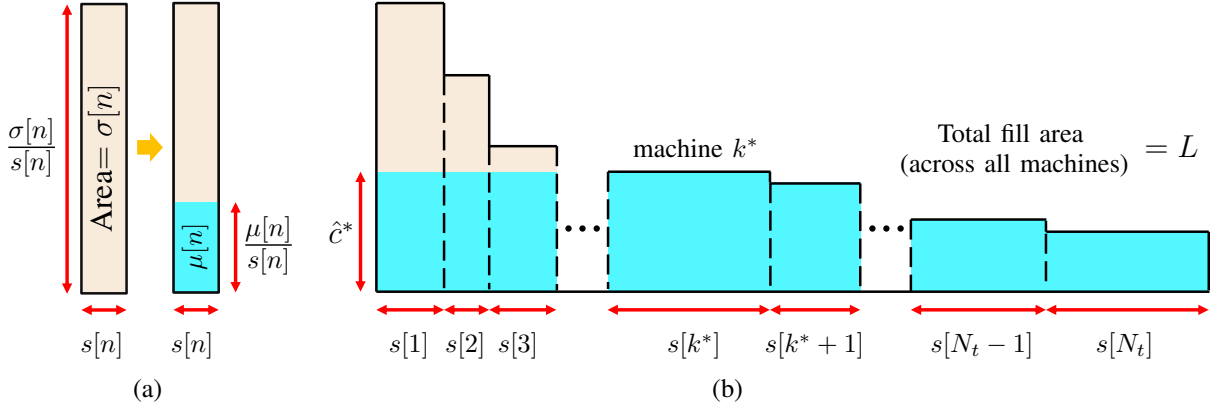

  Finally, we note that the solution in Theorem \ref{th: load_assignment} for the optimization problem of (\ref{eq: optprob}) has a ``water-filling" like visualization as shown in Fig. \ref{fig: waterfill}. 
In Fig. \ref{fig: waterfill}(a), the storage of  machine $n$ is represented by the area of the full rectangle (shaded with the peach color) that it corresponds to. 
We make the width of the rectangle $s[n]$ because if we ``fill'' part of the rectangle with an area of $\mu[n]$ (shaded in blue), then the height of the filled area, which is the water level at that rectangle, represents the computation time of machine $n$. Note that, this filled area does not represent a specific computation assignment, but only the total computations assigned to machine $n$. In Fig. \ref{fig: waterfill}(b), in accordance to (\ref{eq: theorem 1 1}), we arrange  the available machines in descending order of the rectangle height, which is $\frac{\sigma[n]}{s[n]}$ for machine $n$. Then, following (\ref{eq:sum_L}), we ``fill'' all the available  machines with a total area of $L$. First, notice that machines with larger rectangle width, or speed, will have larger filled area, or more computation load, until they are completely filled. Machines $k^*+1,\ldots, N_t$ with smaller rectangle height fill completely and have a computation time strictly less than $\hat{c}^*$. Machines $1,\ldots, k^*$ with larger rectangle height, all have the same computation time of $\hat{c}^*$. Note that one or more of these machines may be completely filled such as machine $k^*$ in Fig. \ref{fig: waterfill}(b), but will have the same ``water level" as machines  $1,\ldots,k^*$.

\subsection{Computation Load of Example 1 and Example 2}
\label{sec: Computation Load Examples}

{\it Homogeneous storage}: We return to Example 1 presented in Section~\ref{sec: Machines With Homogeneous Storage Constraints} with homogeneous storage but  heterogeneous computing speeds and explain how to find the optimal computation load vector. In this example,   each machine stores exactly one  cs-matrix. When $t=1$, we have $N_1=6$ and  $L=3$.
Given $\boldsymbol{s} = [2,\;2,\;3,\;3,\;4,\;4]$, the largest $k^*$ that satisfies (\ref{eq: c_bounds-fork}) is $k^*=6$, and thus $\hat c^*=1/6$,  $\boldsymbol{\mu^*}= \hat c^* \boldsymbol{s}=\left[\frac{1}{3},\frac{1}{3},\frac{1}{2},\frac{1}{2},\frac{2}{3},\frac{2}{3}\right]$.
 Similarly, for $t=2$, since machine 4 is preempted, we have now $N_2=5$, $\Nc_2 = \{1,2,3,5,6\}$ and $\boldsymbol{s} = [2,\;2,\;3,\;4,\;4]$ (we ignore any preempted machines). In this case, we have $k^*=5$, and thus $\hat c^*=1/5$,  $\boldsymbol{\mu^*}= \hat c^* \boldsymbol{s}=\left[\frac{2}{5},\frac{2}{5},\frac{3}{5}, \frac{4}{5},\frac{4}{5}\right].$
Similarly, for $t=3$, we have $N_3=4$, $\Nc_3 = \{1,2,3,5\}$ and $\boldsymbol{s} = [2,\;2,\;3,\;4]$ because machines 4 and 6 preempts.
Here, we have  $k^*=3$, $\hat c^*=2/7$, and $\boldsymbol{\mu^*}=\left[\frac{4}{7},\frac{4}{7},\frac{6}{7}, 1\right]$.
\footnote{Note that, 
as in the optimization problem of (\ref{eq: optprob}), the computation load of the preempted machines are ignored since they are simply $0$, presenting a slight difference between the optimal computation load vectors presented in Section \ref{sec: example}.}

{\it Heterogeneous storage}: We illustrate this case using Example 2 presented in Section~\ref{sec: Machines With Heterogeneous Storage Constraints} with $L=N=6$ and no preempted machines. In this case, we order the machines in a descending order by $\frac{\sigma[n]}{s[n]}$, where $\boldsymbol s = [2, 3, 4, 2, 3, 4]$ and $\boldsymbol \sigma = [2, 2, 2, 1, 1, 1]$. Next, we need to determine $k^*$, and we start by checking $k^*=6$. However, we can observe that  (\ref{eq: c_bounds-fork}) does not hold since
\be
\
\frac{L}{\sum_{n=1}^{N_t}s[n]} 
= \frac{1}{3} > \frac{1}{4} = \frac{\sigma[6]}{s[6]}.
\ee
Similarly, if we try $k^*=5$, we see that (\ref{eq: c_bounds-fork}) does not hold since
\be
\frac{L-\sigma [6]}{\sum_{n=1}^{5}s[n]} = \frac{5}{14}  > \frac{1}{3} = \frac{\sigma[5]}{s[5]}.
\ee
Finally, we see that $k^*=4$ is the solution that satisfies  (\ref{eq: c_bounds-fork}) because
\be
\frac{\sigma[5]}{s[5]} = \frac{1}{3} < \frac{L-\sigma [5] - \sigma [6]}{\sum_{n=1}^{4}s[n]} = \frac{4}{11}  \leq \frac{1}{2} = \frac{\sigma[4]}{s[4]}.
\ee
It follows that  $\hat c^*=4/11$ and by using (\ref{eq:optimal_form}), we obtain $\boldsymbol{\mu^*}=\left[\frac{8}{11},\frac{12}{11},\frac{16}{11}, \frac{8}{11},1,1\right]$.

 In Section \ref{sec: compassign}, we will show that there {always} exists a computation assignment $(\boldsymbol{\mathcal{M}}_t,\boldsymbol{\mathcal{P}}_t)$ whose computation load vector equals $\boldsymbol{\mu^*}$ and the assignment pair can be found using the proposed Algorithm 1 (for homogeneous storage) and Algorithm 2 (for heterogeneous storage) in no more than $N_t$ iterations. 

\section{Second Sub-problem: Optimal Computation Assignment}
\label{sec: compassign}

In this section, we present a computation assignment $(\boldsymbol{\mathcal{M}}_t,\boldsymbol{\mathcal{P}}_t)$ that solves the optimization problem of (\ref{eq: optprob_assign}). First, we show the existence of a computation assignment that yields the computation load vector $\boldsymbol{\mu}^*$ and computation time $\hat{c}^*$. This shows that there is no gap between the combinatorial optimization problem (\ref{eq: optprob_assign}) and the ``relaxed" convex optimization problem of (\ref{eq: optprob}). Then, we provide a low-complexity iterative algorithm that converges to such an assignment in just $N_t$ iterations. In the following, we start with the case of homogeneous storage and heterogeneous computing speeds, where each machine stores exactly one cs-matrix, then we move to the case of heterogeneous storage and computing speed requirements where each machine may store any integer number of cs-matrices.


\subsection{Homogeneous Storage with Heterogeneous Computing Speeds}

Here, we focus on the case where $\sigma[n]=1$ for $n\in[N_t]$ such that each available machine stores exactly one cs-matrix. Our goal is to assign computations among the machines such that each row set in $\boldsymbol{\Mc}_t$ 
is assigned to $L$ machines and the assignments satisfy the $\boldsymbol{\mu}^*$ given by (\ref{eq:optimal_form_hm}). 
{Interestingly, we find that once $\boldsymbol{\mu}^*$ is given, we can adapt the {\em filling problem (FP)} introduced in \cite{woolsey2019optimal} for private information retrieval (PIR) to solve our second sub-problem of finding the computation assignment for CEC networks. Note that our proposed formulation of the computation assignments based on row sets,  together with the two-step approach to solve the proposed combinatorial optimization problem,  are important to allow successful adaptation of the FP problem \cite{woolsey2019optimal} to the CEC setting.  }

In particular, we refer to the following lemma (Theorem~2 in \cite{woolsey2019optimal}).
\begin{lemma}
\label{lemma: 1}
Given $\boldsymbol{\mu}^* \in \mathbb{R}_+^N$ and $L \in \mathbb{Z}^+$, a $(\boldsymbol{\mu}^*, L)$-FP solution exists {\it if and only if}
\begin{align}\label{eq: FP_exist}
\mu^*[n] \leq \frac{\sum_{i=1}^{N_t}\mu^*[i]}{L}
\end{align}
for all $n \in [N_t]$.
\hfill $\square$
\end{lemma}

In our problem setting, we have $\sum_{i=1}^{N_t}\mu^*[i]=L$ and $\mu^*[n]\leq 1$ for all $n \in [N_t]$. Therefore, by using Lemma~\ref{lemma: 1}, an optimal computation assignment exists. Moreover, by adapting Algorithm~1 in \cite{woolsey2019optimal}, we obtain an equivalent Algorithm~\ref{algorithm:1} 
(see  pseudo-codes of Algorithm~\ref{algorithm:1}  for detailed operations) to explicitly provide an optimal computation assignment, $\left(\boldsymbol{\mathcal{M}}_t, \boldsymbol{\mathcal{P}}_t \right)$ where machine $n$ stores and performs computations on its stored cs-matrix $\boldsymbol{\tilde{X}}_n$.

\begin{algorithm}
  \caption{Computation Assignment: Homogeneous Storage Capacity and Heterogeneous Computing Speeds}
  \label{algorithm:1}
  \begin{algorithmic}[1]
  \item[ {\bf Input}: $\boldsymbol{\mu}^*$, $N_t$, $L$, and $q$ ] \quad  $\%$  $\boldsymbol{\mu}^*$ is solution to first sub-problem (\ref{eq:optimal_form})-(\ref{eq: c_bounds-fork}).
  \item $\boldsymbol{m} \leftarrow \boldsymbol{\mu}^*$ \quad  $\%$ $\boldsymbol{m}$ represents the remaining computation load vector to be assigned.  \hspace*{2.6cm} $\%$ Initialize $\boldsymbol{m} $  as the optimal computation load vector $\boldsymbol{\mu}^*$
  \item $f \leftarrow 0$
  \While {$\boldsymbol{m}$ contains a non-zero element}
    \State $f \leftarrow f+1$
    \State $L' \leftarrow \sum_{n=1}^{N_t}m[n]$ \quad \quad $\%$ $L'$ represents the sum of the remaining computation load
    \State $N'\leftarrow$ number of non-zero elements in $\boldsymbol{m}$
    \State $\boldsymbol{\ell} \leftarrow$ indices that sort the non-zero elements of $\boldsymbol{m}$ from smallest to largest\footnotemark[5]
    \State $\mathcal{P}_{f} \leftarrow\{\ell [1], \ell [N'-L+2] , \ldots , \ell [N'] \}$   \quad $\%$ specify  machines that will compute $\mathcal M_f$
    \If {$N' \geq L+1$}  \quad $\%$ $\alpha_f$ is the fraction of rows assigned to row set $\mathcal M_f$
    \State $\alpha_f \leftarrow  \min \left(\frac{L'}{L} - m[\ell[N' - L + 1]], m[\ell[1]]\right)$\footnotemark[6] \quad $\%$ assign only a fraction of remaining  \hspace*{4.5cm} $\%$ rows to to $\mathcal P_f$ to ensure a FP solution exists at next iteration.
    \Else
    \State $\alpha_f \leftarrow  m[\ell[1]]$ \quad $\%$ assign all remaining un-assigned rows of machine $\ell[1]$ to $\mathcal P_f$.
    \EndIf
    \For {$n \in \mathcal{P}_{f}$}
    \State $m[n] \leftarrow m[n] - \alpha_f$ \quad $\%$ update remaining computation load at each machine
    \EndFor
  \EndWhile
  \item $F \leftarrow f$
  \State Partition rows $[\frac{q}{L}]$ into $F$ disjoint row sets: $\mathcal{M}_{1}, \ldots , \mathcal{M}_{F}$ of size $\frac{\alpha_1 q}{L},\ldots,\frac{\alpha_{F}q}{L}$ rows respectively
  \item[ {\bf Output}: $F$, $\mathcal{M}_{1}, \ldots , \mathcal{M}_{F}$ and $\mathcal{P}_{1}, \ldots , \mathcal{P}_{F}$ ]
  \end{algorithmic}
\end{algorithm}

\footnotetext[5]{$\boldsymbol{\ell}$ is an $N'$-length vector and $0<m[\ell[1]]\leq m[\ell[2]]\leq \cdots \leq m[\ell[N']]$.}
\footnotetext[6]{This is the condition obtained by using Lemma~\ref{lemma: 1}.}

\begin{remark}
Using a similar approach in the proof of Lemma~2 in \cite{woolsey2019optimal}, we can show that $F \leq N_t$ such that Algorithm~1 needs at most $N_t$ iterations to complete. Hence, we omit the proof of the correctness of Algorithm~1 here. 
\end{remark}
\begin{remark}
 The connection between Algorithm 1 of this work and that of \cite{woolsey2019optimal} lies in that for the PIR storage placement problem,  one places file sets at $L$ databases one at a time to fulfill certain storage requirement; Analogously, in the second sub-problem of the CEC computation assignment, we allocate computation row sets to $L$ sc-matrices one at a time to fulfill a computation load assignment.
\end{remark}

In the following, we will present an example to illustrate how Algorithm~1 is applied to find the optimal computation assignment. 

\subsection{An Example of Algorithm \ref{algorithm:1} for Homogeneous Storage and Heterogeneous Computing Speed}

\begin{figure}
\centering
\centering \includegraphics[width=8cm]{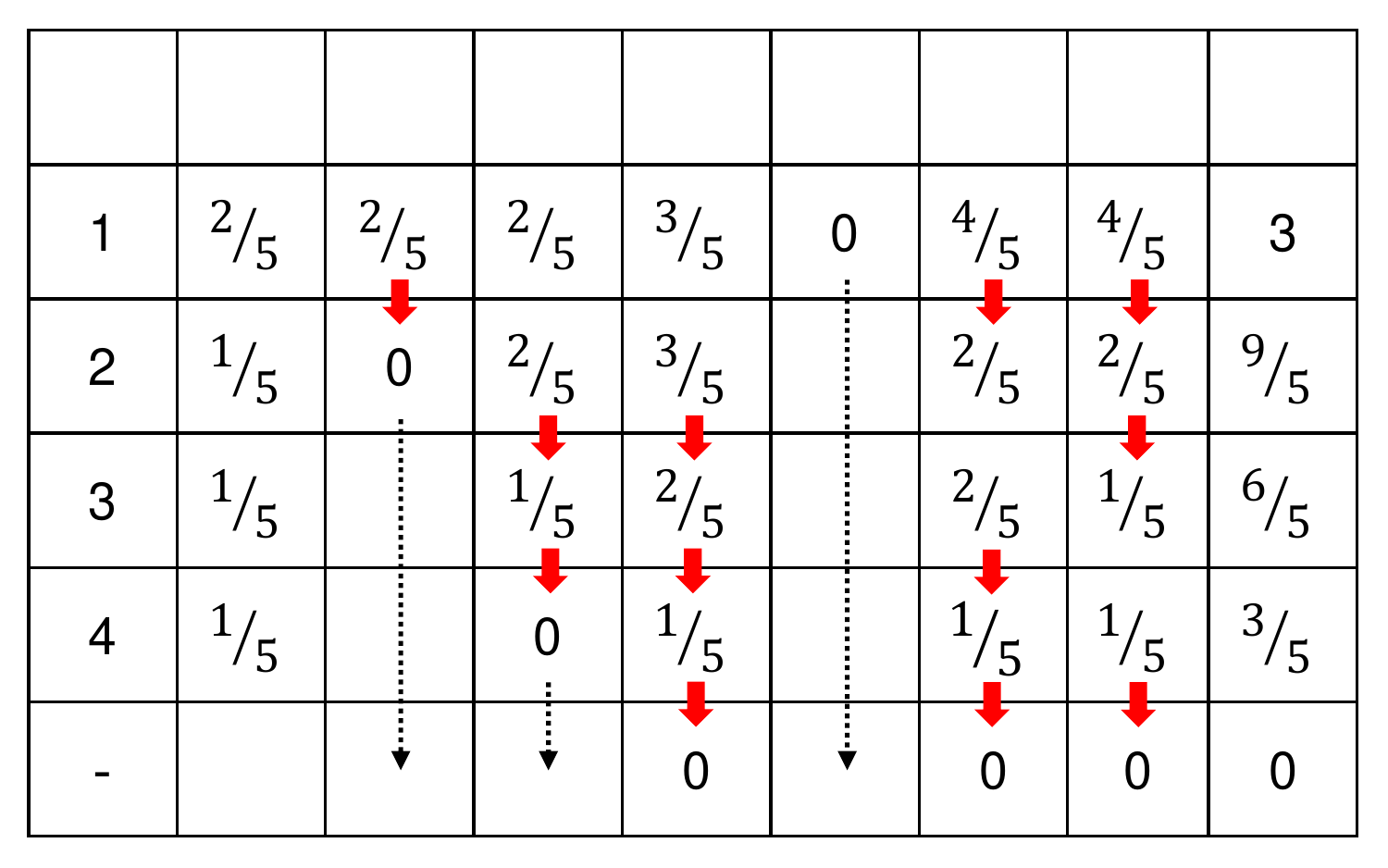} 
\put(-75,43.2){$f$}
\put(-67.5,43.2){$\alpha_f$}
\put(-61,43.2){$m[1]$}
\put(-52.5,43.2){$m[2]$}
\put(-44,43.2){$m[3]$}
\put(-35.5,43.2){$m[4]$}
\put(-27,43.2){$m[5]$}
\put(-18.5,43.2){$m[6]$}
\put(-8,43.2){$L'$}
\vspace{-0.5cm}
\caption{~\small  Computation assignment following Algorithm \ref{algorithm:1} for Example 1 at time $t=2$. Here, $N=6$, $L=3, F=4$. $f$ is the iteration index; Each row corresponds to an iteration $f$, where $\alpha_f$ is to the fraction of rows assigned to row set $\mathcal M_f$; $m[n]$ denotes the remaining computation load for machine $n$ at iteration $f$. The three red arrows from the first row to the second row represent that a fraction of $\alpha_1=\frac{2}{5}$ rows are assigned to $\mathcal M_1$, which are computed by machines (or equivalently, cs-matrices) $\mathcal P_1=\{1,5,6\}$. $L'$ represents the remaining total computation load at iteration $f$. At $f=1$, $L'=L=3$. After one iteration, we have $L'=3-3 \cdot \alpha_1=\frac{9}{5}$. Note that $\alpha_f$ is determined by lines 9 to 12 of Algoirthm \ref{algorithm:1}.
At each iteration $f$,   $\mathcal P_f$ includes the machine with the smallest remaining $m[n]$ and $L-1=2$ machines with the largest remaining $m[n]$.} 
\label{table: example}
\vspace{-0.4cm}
 \end{figure}

We return to Example 1 presented in Sections~\ref{sec: Machines With Homogeneous Storage Constraints} and \ref{sec: Computation Load Examples} 
and use Algorithm \ref{algorithm:1} to derive the computation (rows in each element of $\boldsymbol{\Mc}_t$) assignments for $t=2$, where machine 4 is preempted. The steps of the algorithm are shown in Fig.~\ref{table: example}. In this case, we showed that $\boldsymbol{\mu}^* = \left[\frac{2}{5},\frac{2}{5},\frac{3}{5}, \frac{4}{5},\frac{4}{5}\right]$ in Section~\ref{sec: Computation Load Examples}. In the first iteration, $f=1$, we have $L'=L$ and $\boldsymbol{m}=\boldsymbol{\mu}^*$ as no computations have been assigned yet. Rows of the respective cs-matrices are assigned to machine $1$, $5$, and $6$ because among all  machines, machine $1$ has the least remaining computations to be assigned,  and machines $5$ and $6$ have with the most remaining computations to be assigned. Moreover, note that
\be
\label{eq: m 1}
m[1] = \frac{2}{5} \leq \frac{L'}{L} - m[3] = 1 - \frac{3}{5}=\frac{2}{5},
\ee
where machine $3$ is the machine with the most remaining rows to be assigned that is not included in $\mathcal{P}_1=\{1,5,6\}$. Therefore, a fraction $\alpha_1=\frac{2}{5}$ of the rows are assigned to machines ${1,5,6}$.
Then, $\boldsymbol{m}$ is adjusted to reflect the remaining computations to be assigned and $L'=3-3\alpha_1 = \frac{9}{5}$. For iteration $2$, the condition of (\ref{eq: m 1}) relating to line 10 of Algorithm 1 is motivated by the necessary and sufficient conditions for the existence of a $(\boldsymbol{m}, L')$-FP  solution given in Lemma~\ref{lemma: 1}. Here, we are interested in the FP solution for $\boldsymbol{m}$ which is updated after each iteration. In other words, for each iteration of a row set assignment in Algorithm 1, we ensure there exists a set of succeeding row assignments such that a final FP solution is obtained. 

In the second iteration, $f=2$, machine $2$ is a machine with the least remaining rows to be assigned. Computations are assigned to machine $2$ and machines $3$ and $6$ which are a pair of machines with the most remaining computations to be assigned. Ideally, we would like to assign all the remaining rows to machine $2$. However, since
\be
m[2] = \frac{2}{5} > \frac{L'}{L} - m[5] = \frac{3}{5} - \frac{2}{5}=\frac{1}{5},
\ee
assigning $\frac{2}{5}$ of the rows to machine $2$ in this iteration will violate the condition of (\ref{eq: FP_exist}) in Lemma~\ref{lemma: 1} and as a consequence, there will be no  valid filling solution going forward. Therefore, we set $\alpha_2=\frac{1}{5}$ instead and after this iteration $\boldsymbol{m}$ and $L'$ are adjusted accordingly.

In the third iteration, $f=3$, since
\be
m[2] = \frac{1}{5} \leq \frac{L'}{L} - m[6] = \frac{2}{5} - \frac{1}{5} = \frac{1}{5},
\ee
we assign $\alpha_3=\frac{1}{5}$ of the rows  to machines ${2,3,5}$ and $\boldsymbol{m}$ and $L'$ are adjusted accordingly. Finally, in the fourth iteration, $f=4$, only three machines  ${3,5,6}$ have non-zero computation assignment left and each is assigned  $\alpha_4=\frac{1}{5}$ of the rows. 
In this example, the algorithm converges in $F=4$ (fewer than $N_2=5$) iterations.  The resulting computing assignment is shown in Fig.~\ref{fig: exp1}(b).

\subsection{Heterogeneous Storage Capacity and Computing Speeds}

When both storage capacity and computing speeds among machines are heterogeneous, machines may store more than cs-matrices. In this case, machine $n$ will pick $\lfloor\mu^*[n]\rfloor$ cs-matrices to compute entirely. Then, it will pick the remaining cs-matrix and compute a $\mu^*[n]-\lfloor\mu^*[n]\rfloor$ fraction of that  cs-matrix.  We will show this strategy requires $F\leq N_t$ iterations using Algorithm~2 and the number of computation assignments $F$ is at most  equal to the number of available machines $N_t$. Overall, the assignment consists two steps.
In the first step, those cs-matrices that are computed entirely  are  put into in the cs-matrix sets of $\boldsymbol{\mathcal{P}}_t$. 
In the second step, we assign row sets to the cs-matrices which are not entirely computed so that  each row set in $\boldsymbol \Mc_t$ is guaranteed to be computed across $L$  cs-matrices. Next, we demonstrate that we can  re-use Algorithm \ref{algorithm:1} for the second step of the computation assignment under a modified procedure described in Algorithm~2 (see  pseudo-codes of Algorithm~\ref{algorithm:2}).

To explain the computation assignment process we introduce the following notations. For $n\in[N_t]$, let $\tilde{\mathcal{Q}}_n\subseteq\mathcal{Q}_n$ contain the indices of $\lfloor\mu^*[n]\rfloor$ randomly chosen cs-matrices in $\mathcal{Q}_n$ that machine $n$ computes entirely. Note that  $|\tilde{\mathcal{Q}}_n|=\lfloor\mu^*[n]\rfloor$.  If $\mu[n]<1$, then $\tilde{\mathcal{Q}}_n$ is empty. Next, machine $n$ randomly chooses one cs-matrix from 
$\mathcal{Q}_n\setminus \tilde{\mathcal{Q}}_n$ to compute partially. 
Note that when  $\mu[n]$ is an integer,  $\hat{\theta}[n]$ is simply a dummy variable and is never referenced, i.e. $\hat{\theta}[n] = \varnothing$. In the following, we denote $\hat{\boldsymbol{\theta}} = [\hat{\theta}[1], \hat{\theta}[2], \ldots, \hat{\theta}[N_t]]$. Then we define the partial computation vector $\hat{\boldsymbol{\mu}}\in\mathbb{R}_+^{N_t}$ such that
\be
\label{eq: mu partial}
\hat{\mu}[n] = \mu^*[n]-\lfloor\mu^*[n]\rfloor, \;\; \forall n \in [N_t].
\ee
Hence, machine $n$ will entirely compute each cs-matrix $\boldsymbol{\tilde{X}}_i$ for $i\in\tilde{\mathcal{Q}}_n$ and compute a  $\hat{\mu}[n]$ fraction of the cs-matrix $\boldsymbol{\tilde{X}}_{\hat{\theta}[n]}$.

\begin{algorithm}
  \caption{ Computation Assignment: Heterogeneous Storage Capacity and Computing Speeds}
  \label{algorithm:2}
  \begin{algorithmic}[1]
  \item[ {\bf Input}: $\boldsymbol{\mu}^*$, $N_t$, $L$, $q$, $\tilde{\mathcal{Q}}_1,\ldots,\tilde{\mathcal{Q}}_{N_t}$ and $\hat{\boldsymbol{\theta}}$ ]
   \quad  $\%$  $\boldsymbol{\mu}^*$ is solution to first sub-problem (\ref{eq:optimal_form})-(\ref{eq: c_bounds-fork}).
   \hspace*{4cm} $\%$ $\tilde{\mathcal{Q}}_1,\ldots,\tilde{\mathcal{Q}}_{N_t}$ and $\hat{\boldsymbol{\theta}}$ include pre-chosen cs-matrices based on   $\boldsymbol{\mu}^*$.
  \For {$n \in [N_t]$}
  \State $\hat{\mu}[n] \leftarrow \mu^*[n]-\lfloor\mu^*[n]\rfloor$ \quad $\%$ determine the computation load for each partially computed \hspace*{5cm} $\%$cs-matrix. 
  \EndFor
  \item $\hat{L} \leftarrow \sum_{n=1}^{N_t}\hat{\mu}[n]$ \quad \quad $\%$ $\hat L$ is the sum of the computation load over partially computed  \hspace*{3.8cm} $\%$ cs-matrices. Use Algorithm 1 next to find computation assignment  \hspace*{3.8cm} $\%$ for partially computed matrices. 
  \item $\hat{F}$, $\hat{\mathcal{M}}_1,\ldots,\hat{\mathcal{M}}_F$ and $\hat{\mathcal{P}}_1,\ldots,\hat{\mathcal{P}}_F$ $\leftarrow$ Output of {\bf Algorithm \ref{algorithm:1}} with $\hat{\boldsymbol{\mu}}$, $N_t$, $\hat{L}$, and $q$ as input
  \item $F \leftarrow \hat{F}$
  \For {$f \in [F]$}
  \State $\mathcal{M}_f\leftarrow \hat{\mathcal{M}}_f$ \hspace*{1.5cm} $\%$ use the output row sets from Algorithm 1 as final row sets.
  \State $\mathcal{P}_f\leftarrow \bigcup_{n\in\hat{\mathcal{P}}_f} \hat{\theta}[n] \cup \bigcup_{n\in[N_t]} \tilde{\mathcal{Q}}_n$ \quad $\%$ combine outputs of Algorithm 1 with fully computed \hspace*{6.2cm} $\%$ cs-matrices to obtain the final cs-matrices assignment 
  \EndFor
  \item[ {\bf Output}: $F$, $\mathcal{M}_{1}, \ldots , \mathcal{M}_{F}$ and $\mathcal{P}_{1}, \ldots , \mathcal{P}_{F}$ ]
  \end{algorithmic}
\end{algorithm}

Finally, we define the sum of the partial computation load vector $\hat{\boldsymbol{\mu}}$ as
\be
\hat{L} \triangleq \sum_{n=1}^{N_t}\hat{\mu}[n].
\ee
Note that $\hat{L}$ is an important parameter because it represents the number of cs-matrices that each row set needs to be assigned to, excluding those  cs-matrices that are entirely computed. In other words, since the elements in $\boldsymbol{\mu}$ sum to $L$, there are $L-\hat{L}$ cs-matrices that are entirely computed by the machines. Therefore, in order to assign each row computation (or row set) in $\boldsymbol \Mc_t$ to $L$ cs-matrices, we assign each row set to $\hat{L}$ cs-matrices that are only partially computed. The detailed description of the proposed algorithm is given in Algorithm~2.

In Algorithm~2, we will perform Algorithm \ref{algorithm:1} on $\hat{\boldsymbol{\mu}}$, which iteratively fills some computations at $\hat{L}$ machines in each iteration. Following similar arguments as before, we need to ensure that such a filling problem solution exists and  can be found using the proposed algorithm. 
From Lemma~\ref{lemma: 1}, we see that a $(\hat{\boldsymbol{\mu}}, \hat L)$-FP solution exists because $\hat{\mu}[n] < 1$ for all $n\in[N_t]$ and $\hat{L} = \sum_{n=1}^{N_t}\hat{\mu}[n]$. 
Then, similar to the previous analysis, it can be seen that Algorithm \ref{algorithm:1} will yield a $(\hat{\boldsymbol{\mu}}, \hat L)$-FP solution with $F\leq N_t$ iterations.
This means that, in order to use Algorithm \ref{algorithm:1}, instead of inputting $\boldsymbol{\mu}^*$ and $L$, we input $\hat{\boldsymbol{\mu}}$ and $\hat{L}$. Then, we label the output of Algorithm \ref{algorithm:1} as $\hat{F}$, $\hat{\mathcal{M}}_1,\ldots,\hat{\mathcal{M}}_F$ and $\hat{\mathcal{P}}_1,\ldots,\hat{\mathcal{P}}_F$. These variables represent the computation assignments at the cs-matrices that are partially computed, but computations are only assigned to $\hat{L}$, instead of $L$, cs-matrices. Note that, due to the one-to-one mapping between partially computed cs-matrices and machines, each $\hat{\mathcal{P}}_f$ represents the set of machines that are assigned to compute rows in $\mathcal{M}_f$ of the cs-matrix it is partially computing.
To complete the computation assignment, we must include the $L-\hat{L}$ cs-matrices that are entirely computed. Therefore,
\be
\mathcal{P}_f = \bigcup_{n\in\hat{\mathcal{P}}_f} \hat{\theta}[n] \; \cup  \bigcup_{m\in[N_t]} \tilde{\mathcal{Q}}_m , \;\; \forall f\in[F],
\ee
where $|\bigcup_{n\in\hat{\mathcal{P}}_f} \hat{\theta}[n]| = \hat L$ and $| \bigcup_{m\in[N_t]} \tilde{\mathcal{Q}}_m| = L-\hat L$.

Since Algorithm \ref{algorithm:1} assigns computations to machines\footnote{Algorithm \ref{algorithm:1} is designed with the assumption that machine $n, \forall n \in [N]$ stores one cs-matrix, $\tilde{X}_n$.},   $\hat{\boldsymbol{\theta}}$ is needed to identify the  cs-matrices that the machines partially compute. In addition, the number of computation assignments remains the same and $\hat{F}=F$. The row sets also remain the same, $\mathcal{M}_f=\hat{\mathcal{M}}_f$ for all $f\in[F]$.

Algorithm~2 will be illustrated using the following example.


\subsection{An Example of Algorithm \ref{algorithm:2} for Heterogeneous Storage and Computing Speed}

We consider Example 2 presented in Sections~\ref{sec: Machines With Heterogeneous Storage Constraints} and  \ref{sec: Computation Load Examples}, where we have $L=6$ and $\boldsymbol \sigma = [2, 2, 2, 1, 1, 1]$. This means that machines $1$, $2$, and $3$ stores two cs-matrices while machines $4$, $5$, and $6$ stores one cs-matrix, respectively. 
We assume no preempted machines at $t=1$. In this case, the optimal computation load vector is found to be $\boldsymbol{\mu}^* = \left[\;\frac{8}{11},\;\;\frac{12}{11},\;\;\frac{16}{11},\;\;\frac{8}{11},\;\;1,\;\;1\;\right]$ in Section~\ref{sec: Computation Load Examples}. Since $\mu^*[n] \geq 1, n \in \{2, 3, 5, 6\}$, it can be seen that machines $2$, $3$, $5$, and $6$ will compute all the rows sets of $\boldsymbol\Mc_1$ for one cs-matrix (see Fig.~\ref{fig: exp2}). Next, machines $1$, $2$, $3$ and $4$ have one cs-matrix to be partially computed, and each of them will compute a fraction of that cs-matrix.
Note that, since $\mu^*[5]=1$ and $\mu^*[6]=1$ are integers, by the algorithm design, no computations will be assigned to cs-matrices partially computed by machines $5$ and $6$. In other words, based on the optimal computation load vector $\boldsymbol{\mu}^*$, machines $5$ and $6$ only entirely compute  cs-matrices. Using (\ref{eq: mu partial}), we can obtain the partial computation load vector as 
$\hat{\boldsymbol{\mu}} = \left[\;\frac{8}{11},\;\;\frac{1}{11},\;\;\frac{5}{11},\;\;\frac{8}{11},\;\;0,\;\;0\;\right]$,
whose elements sum to $\hat{L}=2$. Our goal is to assign computations to cs-matrices partially computed by machines $1$ through $4$, where we assign the computations corresponding to each row set of $\boldsymbol\Mc_1$ to $\hat{L}=2$  cs-matrices at a time. 
This will be done using Algorithm \ref{algorithm:1}.

In particular, let the indexes of the cs-matrices stored at the machines be $\mathcal{Q}_1=\{1,2 \}$, $\mathcal{Q}_2=\{3,4 \}$, $\mathcal{Q}_3=\{5,6 \}$, $\mathcal{Q}_4=\{7 \}$, $\mathcal{Q}_5=\{8 \}$ and $\mathcal{Q}_6=\{9 \}$. Each machine $n$ picks a set of $\lfloor \mu[n]\rfloor$ stored cs-matrices to be computed entirely which 
could be that of $\tilde{\mathcal{Q}}_1=\varnothing$, $\tilde{\mathcal{Q}}_2=\{3 \}$, $\tilde{\mathcal{Q}}_3=\{5 \}$, $\tilde{\mathcal{Q}}_4=\varnothing$, $\tilde{\mathcal{Q}}_5=\{8 \}$ and $\tilde{\mathcal{Q}}_6=\{9 \}$. Moreover, each machine selects an index of a stored cs-matrix to be partially computed, which are denoted as 
$\hat{\boldsymbol{\theta}}=\left[\;1,\;\;4,\;\;6,\;\;7,\;\;0,\;\;0\;\right]$.

In the first iteration of Algorithm \ref{algorithm:1} inside Algorithm 2 (line 5), we aim to assign some computations to the cs-matrix $\boldsymbol{\tilde{X}}_{\hat{\theta}[2]}$, since $\hat \mu[2]$ is the smallest non-zero element in $\hat{\boldsymbol{\mu}}$. 
$\boldsymbol{\tilde{X}}_{\hat{\theta}[2]}$ will be partially computed by machine $2$. We also assign this computation to machine $4$ because it is a machine with the largest remaining computations to be assigned (line 8 in Algorithm 1). Therefore, we assign a $\alpha_1=\frac{1}{11}$ fraction of rows to the cs-matrices partially computed by machines $2$ and $4$  (line 10 in Algorithm 1). After this iteration $\boldsymbol{m}=\left[\;\frac{8}{11},\;\;0,\;\;\frac{5}{11},\;\;\frac{7}{11},\;\;0,\;\;0\;\right]$  (line 15 in Algorithm 1), and machines $1$ and $3$ are the machines with the most and least, respectively, remaining computations to be assigned. From line 10 in Algorithm~1, since
\be
m[3]=\frac{5}{11}> \frac{3}{11}=\frac{\hat{L}'}{\hat{L}}-m[4],
\ee
we assign a $\alpha_2=\frac{3}{11}$ fraction of rows to machines $1$ and $3$. Then, 
after this iteration,
we find $\boldsymbol{m} = \left[ \; \frac{5}{11}, \;\;0,\;\;\frac{2}{11},\;\;\frac{7}{11},\;\;0,\;\;0\;\right]$. By a similar approach, next we assign a $\alpha_3=\frac{2}{11}$ fraction of rows to machines $3$ and $4$ and a $\alpha_4=\frac{5}{11}$ fraction of rows to machines $1$ and $4$. After this iteration, we can find that $\boldsymbol{m}=\boldsymbol{0}$.

Based on above procedure in  Algorithm \ref{algorithm:1}, we obtain the output $\hat{F}$, $\hat{\mathcal{M}}_1,\ldots,\hat{\mathcal{M}}_4$ which contain a $\frac{1}{11}$, $\frac{3}{11}$, $\frac{2}{11}$ and $\frac{5}{11}$ fraction of rows, respectively, and the machines assigned to compute row sets $\mathcal M_f$ are given by $\hat{\mathcal{P}}_1=\{ 2,4 \}$, $\hat{\mathcal{P}}_2=\{ 1,3 \}$, $\hat{\mathcal{P}}_3=\{ 3,4 \}$ and $\hat{\mathcal{P}}_4=\{ 1,4 \}$.
For the final solution, the number of assignments stays the same, $F=\hat{F}=4$, and the row sets stay the same $\mathcal{M}_f=\hat{\mathcal{M}}_f,\;\forall f\in [F]$. However, using $\hat{\mathcal{P}}_f,\;\forall f\in [F]$, we need to define specifically which cs-matrices are being computed for each row set. Note that, $\hat{\mathcal{P}}_f$ is the set of machines that are assigned to compute rows in $\mathcal{M}_f$ of the corresponding cs-matrix partially computed. We use $\hat{\boldsymbol{\theta}}$ to resolve the indexes of the cs-matrices from $\hat{\mathcal{P}}_f$. Then, we also need to include the indexes of all cs-matrices that are entirely computed from $\tilde{\mathcal{Q}}_1,\ldots ,\tilde{\mathcal{Q}}_6$. For example, recall that, $\tilde{\mathcal{Q}}_1=\varnothing$, $\tilde{\mathcal{Q}}_2=\{3 \}$, $\tilde{\mathcal{Q}}_3=\{5 \}$, $\tilde{\mathcal{Q}}_4=\varnothing$, $\tilde{\mathcal{Q}}_5=\{8 \}$ and $\tilde{\mathcal{Q}}_6=\{9 \}$, we then obtain the cs-matrix sets
\be
\mathcal{P}_1=\{\hat{\theta}[2], \hat{\theta}[4],3,5,8,9 \}=\{3,4,5,7,8,9\}.
\ee
Similarly, we see that $\mathcal{P}_2=\{1,3,5,6,8,9\}$, $\mathcal{P}_3=\{3,5,6,7,8,9\}$ and $\mathcal{P}_4=\{1,3,5,7,8,9\}$.

\section{Conclusions}
\label{sec: conclusions}
In this paper, we study the heterogeneous coded elastic computing problem where computing machines store MDS coded data matrices and may have both varying computation speeds and storage capacity. The key of this problem is to design a fixed storage assignment scheme and a computation assignment strategy such that no redundant computations are present and the overall computation time can be minimized as long as there {are at least $L$ cs-matrices stored among the available machines}.  Given a set of available machines with arbitrary relative computation speeds and storage capacity, we first proposed a novel combinatorial min-max problem formulation in order to minimize the overall computation time, which is determined by the machines that need the longest computation time. Based on the MDS coded storage assignment, the goal of this optimization problem is to assign computation tasks to machines such that the overall computation time is minimized.
In order to precisely solve this combinatorial problem, we decompose it into a convex optimization problem to determine the optimal computation load of each machine and a computation assignment problem {that yields} the resulting computation load from the convex optimization problem. Then, we adapt low-complexity iterative algorithms to find the optimal solution to the original combinatorial problems, which require a number of iterations no greater than the number of available machines. The proposed heterogeneous coded elastic computing design has the potential to perform computations faster than the state-of-the-art design which was developed for a homogeneous distributed computing system.


\appendices

\bibliographystyle{IEEEbib}
\bibliography{references_d2d}

\end{document}

%% file: macros.tex
\newfont{\bbb}{msbm10 scaled 700}

\newfont{\bb}{msbm10 scaled 1100}

\newcommand{\Mc}{{\cal M}}
\newcommand{\Nc}{{\cal N}}

\newcommand{\Pc}{{\cal P}}
\newcommand{\Qc}{{\cal Q}}

\newcommand{\be}{\begin{equation}}
\newcommand{\ee}{\end{equation}}
\newcommand{\bea}{\begin{eqnarray}}
\newcommand{\eea}{\end{eqnarray}}



%% file: CEC_HetStorage_v5.bbl
\begin{thebibliography}{10}

\bibitem{woolsey2019hetCEC}
N.~Woolsey, R.~Chen, and M.~Ji,
\newblock ``Heterogeneous computation assignments in coded elastic computing,''
\newblock {\em arXiv preprint arXiv:2001.04005}, 2020.

\bibitem{li2018fundamental}
S.~Li, M.~A. Maddah-Ali, Q.~Yu, and A.~S. Avestimehr,
\newblock ``A fundamental tradeoff between computation and communication in
  distributed computing,''
\newblock {\em IEEE Transactions on Information Theory}, vol. 64, no. 1, pp.
  109--128, 2018.

\bibitem{li2018cdc}
S.~{Li}, M.~A. {Maddah-Ali}, and A.~S. {Avestimehr},
\newblock ``Compressed coded distributed computing,''
\newblock in {\em 2018 IEEE International Symposium on Information Theory
  (ISIT)}, 2018, pp. 2032--2036.

\bibitem{konstantinidis2018leveraging}
K.~{Konstantinidis} and A.~{Ramamoorthy},
\newblock ``Resolvable designs for speeding up distributed computing,''
\newblock {\em IEEE/ACM Transactions on Networking}, pp. 1--14, 2020.

\bibitem{woolsey2018new}
N.~Woolsey, R.~Chen, and M.~Ji,
\newblock ``A new combinatorial design of coded distributed computing,''
\newblock in {\em 2018 IEEE International Symposium on Information Theory
  (ISIT)}. IEEE, 2018, pp. 726--730.

\bibitem{Srinivasavaradhan2018distributed}
S.~R. Srinivasavaradhan, L.~Song, and C.~Fragouli,
\newblock ``Distributed computing trade-offs with random connectivity,''
\newblock in {\em Proc. IEEE Int. Symp. Inf. Theory}, 2018.

\bibitem{prakash2018coded}
S.~Prakash, A.~Reisizadeh, R.~Pedarsani, and S.~Avestimehr,
\newblock ``Coded computing for distributed graph analytics,''
\newblock {\em arXiv:1801.05522}, 2018.

\bibitem{woolsey2019ccdc}
N.~{Woolsey}, R.~{Chen}, and M.~{Ji},
\newblock ``Cascaded coded distributed computing on heterogeneous networks,''
\newblock in {\em 2019 IEEE International Symposium on Information Theory
  (ISIT)}, 2019, pp. 2644--2648.

\bibitem{xu2019cdc}
F.~{Xu} and M.~{Tao},
\newblock ``Heterogeneous coded distributed computing: Joint design of file
  allocation and function assignment,''
\newblock in {\em 2019 IEEE Global Communications Conference (GLOBECOM)}, 2019,
  pp. 1--6.

\bibitem{woolsey2019coded}
N.~Woolsey, R.-R. Chen, and M.~Ji,
\newblock ``Coded distributed computing with heterogeneous function
  assignments,''
\newblock {\em arXiv preprint arXiv:1902.10738}, 2019.

\bibitem{wan2020topological}
K.~Wan, M.~Ji, and G.~Caire,
\newblock ``Topological coded distributed computing,''
\newblock {\em arXiv preprint arXiv:2004.04421}, 2020.

\bibitem{attia2019shuffling}
M.~{Adel Attia} and R.~{Tandon},
\newblock ``Near optimal coded data shuffling for distributed learning,''
\newblock {\em IEEE Transactions on Information Theory}, vol. 65, no. 11, pp.
  7325--7349, Nov 2019.

\bibitem{elmahdy2018shuffling}
A.~{Elmahdy} and S.~{Mohajer},
\newblock ``On the fundamental limits of coded data shuffling,''
\newblock in {\em 2018 IEEE International Symposium on Information Theory
  (ISIT)}, June 2018, pp. 716--720.

\bibitem{wan2020fundamental}
K.~Wan, D.~Tuninetti, M.~Ji, G.~Caire, and P.~Piantanida,
\newblock ``Fundamental limits of decentralized data shuffling,''
\newblock {\em IEEE Transactions on Information Theory}, 2020.

\bibitem{lee2017speeding}
K.~Lee, M.~Lam, R.~Pedarsani, D.~Papailiopoulos, and K.~Ramchandran,
\newblock ``Speeding up distributed machine learning using codes,''
\newblock {\em IEEE Transactions on Information Theory}, vol. PP, no. 99, pp.
  1--1, 2017.

\bibitem{tandon2017gradient}
R.~Tandon, Qi~Lei, A.~G. Dimakis, and N.~Karampatziakis,
\newblock ``Gradient coding: Avoiding stragglers in distributed learning,''
\newblock in {\em International Conference on Machine Learning}, 2017, pp.
  3368--3376.

\bibitem{dutta2016short}
S.~Dutta, V.~Cadambe, and P.~Grover,
\newblock ``Short-dot: Computing large linear transforms distributedly using
  coded short dot products,''
\newblock in {\em Advances In Neural Information Processing Systems}, 2016, pp.
  2100--2108.

\bibitem{wan2020distributed}
K.~Wan, H.~Sun, M.~Ji, and G.~Caire,
\newblock ``Distributed linearly separable computation,''
\newblock {\em arXiv preprint arXiv:2007.00345}, 2020.

\bibitem{yu2020straggler}
Q.~Yu, M.~A. Maddah-Ali, and A.~S. Avestimehr,
\newblock ``Straggler mitigation in distributed matrix multiplication:
  Fundamental limits and optimal coding,''
\newblock {\em IEEE Transactions on Information Theory}, vol. 66, no. 3, pp.
  1920--1933, 2020.

\bibitem{dutta2020coding}
S.~{Dutta}, M.~{Fahim}, F.~{Haddadpour}, H.~{Jeong}, V.~{Cadambe}, and
  P.~{Grover},
\newblock ``On the optimal recovery threshold of coded matrix multiplication,''
\newblock {\em IEEE Transactions on Information Theory}, vol. 66, no. 1, pp.
  278--301, 2020.

\bibitem{wan2020cache}
K.~Wan, H.~Sun, M.~Ji, D.~Tuninetti, and G.~Caire,
\newblock ``Cache-aided matrix multiplication retrieval,''
\newblock {\em arXiv preprint arXiv:2007.00856}, 2020.

\bibitem{karakus2017straggler}
C.~Karakus, Y.~Sun, S.~Diggavi, and W.~Yin,
\newblock ``Straggler mitigation in distributed optimization through data
  encoding,''
\newblock in {\em Advances in Neural Information Processing Systems}, 2017, pp.
  5434--5442.

\bibitem{bitar2017minimizing}
R.~Bitar, P.~Parag, and S.~El~Rouayheb,
\newblock ``Minimizing latency for secure distributed computing,''
\newblock in {\em Information Theory (ISIT), 2017 IEEE International Symposium
  on}. IEEE, 2017, pp. 2900--2904.

\bibitem{aliasgari2020computing}
M.~{Aliasgari}, O.~{Simeone}, and J.~{Kliewer},
\newblock ``Private and secure distributed matrix multiplication with flexible
  communication load,''
\newblock {\em IEEE Transactions on Information Forensics and Security}, vol.
  15, pp. 2722--2734, 2020.

\bibitem{hua2019privacy}
H.~{Sun} and S.~A. {Jafar},
\newblock ``The capacity of private computation,''
\newblock {\em IEEE Transactions on Information Theory}, vol. 65, no. 6, pp.
  3880--3897, 2019.

\bibitem{obead2020private}
S.~A. Obead, H.-Y. Lin, E.~Rosnes, and J.~Kliewer,
\newblock ``Private function computation for noncolluding coded databases,''
\newblock {\em arXiv preprint arXiv:2003.10007}, 2020.

\bibitem{chen2020gcsa}
Z.~Chen, Z.~Jia, Z.~Wang, and S.~A. Jafar,
\newblock ``Gcsa codes with noise alignment for secure coded multi-party batch
  matrix multiplication,''
\newblock {\em arXiv preprint arXiv:2002.07750}, 2020.

\bibitem{chen2018draco}
L.~Chen, H.~Wang, Z.~Charles, and D.~Papailiopoulos,
\newblock ``Draco: Byzantine-resilient distributed training via redundant
  gradients,''
\newblock in {\em International Conference on Machine Learning}, 2018, pp.
  902--911.

\bibitem{jonas2017occupy}
E.~Jonas, Q.~Pu, S.~Venkataraman, I.~Stoica, and B.~Recht,
\newblock ``Occupy the cloud: Distributed computing for the 99\%,''
\newblock in {\em Proceedings of the 2017 Symposium on Cloud Computing}, 2017,
  pp. 445--451.

\bibitem{yang2018coded}
Y.~{Yang}, M.~{Interlandi}, P.~{Grover}, S.~{Kar}, S.~{Amizadeh}, and
  M.~{Weimer},
\newblock ``Coded elastic computing,''
\newblock in {\em 2019 IEEE International Symposium on Information Theory
  (ISIT)}, July 2019, pp. 2654--2658.

\bibitem{dau2019optimizing}
H.~Dau, R.~Gabrys, Y.~Huang, C.~Feng, Q.~Luu, E.~Alzahrani, and Z.~Tari,
\newblock ``Optimizing the transition waste in coded elastic computing,''
\newblock {\em arXiv preprint arXiv:1910.00796}, 2019.

\bibitem{woolsey2019optimal}
N.~Woolsey, R.~Chen, and M.~Ji,
\newblock ``An optimal iterative placement algorithm for pir from heterogeneous
  storage-constrained databases,''
\newblock in {\em GLOBECOM 2019 IEEE Global Communications Conference}. IEEE,
  2019.

\bibitem{yang2018codedArxiv}
Y.~{Yang}, M.~{Interlandi}, P.~{Grover}, S.~{Kar}, S.~{Amizadeh}, and
  M.~{Weimer},
\newblock ``Coded elastic computing,''
\newblock {\em arXiv preprint arXiv:1812.06411}, 2018.

\end{thebibliography}
